\documentclass[11pt]{article}
\pdfoutput=1

\usepackage{booktabs}
\usepackage[english]{babel}
\usepackage{amsmath,amssymb,amsbsy,amstext, amsthm, simplewick}
\usepackage{hyperref}
\usepackage{graphicx}
\usepackage{amsfonts}
\usepackage{amssymb}
\usepackage{upgreek}
\usepackage{simplewick}
 \usepackage{exscale,relsize}
\usepackage{mathtools}
\usepackage{comment}

\usepackage[margin=1cm,labelfont={sf,bf,scriptsize},textfont={sf,scriptsize}]{caption}

\usepackage{colortbl}
\definecolor{lightgreen}{cmyk}{0.2, 0, 0.2, 0.2}
\definecolor{lightgray}{cmyk}{0.1,0.2,0,0.1}
\definecolor{lightgray2}{cmyk}{0.1,0.1,0,0.1}

\makeatletter
\newlength{\apb@width}
\newcommand{\autoparbox}[2][c]{\settowidth{\apb@width}{#2}\parbox[#1]{\apb@width}{#2}}

\makeatother

\numberwithin{equation}{section}

\def\beq{\begin{equation}}
\def\eeq{\end{equation}}

\def\bea{\begin{eqnarray}}
\def\eea{\end{eqnarray}}

\def\d{{\rm d}}

\def\beq{\begin{equation}}
\def\eeq{\end{equation}}
\def\be{\begin{equation}}
\def\ee{\end{equation}}
\def\bea{\begin{eqnarray}}
\def\eea{\end{eqnarray}}

\def\d{{\rm d}}

\def\d{{\rm d}}

\def\0{{\vec{0}}}

\DeclareRobustCommand{\SkipTocEntry}[4]{}

\def\d{{\rm d}}

\def\beq{\begin{equation}}
\def\eeq{\end{equation}}
\def\d{\partial}

\def\d{\partial}

\def\ba#1\ea{\begin{align}#1\end{align}}
\def\bg#1\eg{\begin{gather}#1\end{gather}}
\newcommand{\bseq}{\begin{subequations}}
\newcommand{\eseq}{\end{subequations}}

\renewcommand{\t}{\tilde}
\newcommand{\mR}{{\mathcal R}}

\DeclareSymbolFont{extraup}{U}{zavm}{m}{n}
\DeclareMathSymbol{\varheart}{\mathalpha}{extraup}{86}
\DeclareMathSymbol{\vardiamond}{\mathalpha}{extraup}{87}

\def\({\left(}
\def\){\right)}
\def\[{\left[}
\def\]{\right]}
\def\eff{{\rm eff}}

\begin{document}

\begin{titlepage}

\setcounter{page}{1} \baselineskip=15.5pt \thispagestyle{empty}

\vbox{\baselineskip14pt
}
{~~~~~~~~~~~~~~~~~~~~~~~~~~~~~~~~~~~~
~~~~~~~~~~~~~~~~~~~~~~~~~~~~~~~~~~
~~~~~~~~~~~ \footnotesize{SLAC-PUB-15800, SU/ITP-13/20}} \date{}

\bigskip\

\vspace{2cm}
\begin{center}
{\fontsize{19}{36}\selectfont  \sc
New solutions with accelerated expansion\\ 
\vspace{5mm}
in string theory
}
\end{center}

\vspace{0.6cm}

\begin{center}
{\fontsize{13}{30}\selectfont  Matthew Dodelson$^1$, Xi Dong$^1$, Eva Silverstein$^{1,2,3}$, and Gonzalo Torroba$^1$}
\end{center}


\begin{center}
\vskip 8pt
\textsl{
$^1$Stanford Institute for Theoretical Physics, Stanford University, Stanford, CA 94306, USA}

\vskip 7pt
\textsl{ $^2$SLAC National Accelerator Laboratory, 2575 Sand Hill, Menlo Park, CA 94025}

\vskip 7pt
\textsl{ $^3$Kavli Institute for Particle Astrophysics and Cosmology, Stanford, CA 94025, USA}


\end{center}

\vspace{1.2cm}
\hrule \vspace{0.3cm}
{ \noindent \textbf{Abstract} \\[0.2cm]
\noindent 
We present concrete solutions with accelerated expansion in string theory, requiring a small, tractable list of stress energy sources.  We explain how this construction (and others in progress) evades previous no go theorems for simple accelerating solutions.  Our solutions respect an approximate scaling symmetry and realize discrete sequences of values for the equation of state, including one with an accumulation point at $w=-1$ and another accumulating near $w=-1/3$ from below.  In another class of models, a density of defects generates scaling solutions with accelerated expansion.  We briefly discuss potential applications to dark energy phenomenology, and to holography for cosmology.  
}

 \vspace{0.3cm}
 \hrule

\vspace{0.6cm}
\end{titlepage}

\tableofcontents

\newpage
\section{Introduction and summary}

Much of the progress that has occurred in string theory and its applications exploits explicit and tractable background solutions.  In the most generic, and realistic, setting of time-dependent solutions and cosmology, although much has been learned \cite{landscapereviews, inflationreviews}\ it is fair to say that the most realistic backgrounds with metastabilized moduli that have been studied such as \cite{SCdS,KKLT,large, Saltman}\ are relatively complicated.\footnote{Nonetheless simple physical mechanisms have been discovered within these systems which are tractable and phenomenologically useful in themselves \cite{inflationreviews}.}  Especially when it comes to conceptual questions about how to formulate cosmological observables, interpret horizon entropy, and resolve singularities, explicit examples would seem to be particularly useful.  In some specific classes of solutions, concrete lessons have emerged about generalizations of string dualities (see e.g. \cite{DimMutation,simeon}) and holography (see e.g. \cite{dSCFT, dSdS, micromanaging}), but this is just the tip of what promises to be a much larger iceberg.  

In this paper, we revisit the problem of deriving simple solutions realizing accelerated expansion in string theory, finding an explicit set of tractable models with a discrete distribution of accelerating equations of state and an approximate scaling symmetry.  Rather than embedding inflation or quintessence into a separate moduli stabilization scenario, we incorporate the moduli more directly into a rolling scalar solution, making use of various inflationary mechanisms developed over the years.   The leading term in the perturbative string moduli potential, and the axions that dominate the string spectrum, play an essential role in our first class of solutions.  Another set makes use of a leading source of domain walls to obtain accelerated expansion, with the solution taking into account the moduli-dependence of their tension.    
   
The resulting sequences of possible equations of state, some of which accumulate near $w=-1$, may have application to dark energy (for another recent string theoretic example see \cite{SandipDE}).  This variety of equations of state is also interesting for more formal applications -- along with the scaling behavior it enters into the formulas for the entropy and other aspects of a putative holographic dual description (such as the evolution of couplings  on the branes that describe the approximate Coulomb branch of the dual).  For all the accelerating equations of state, there is a future horizon and future spacelike infinity similarly to pure de Sitter spacetimes \cite{HellSuss}, with a gravitational entropy whose microscopic interpretation is of interest.   Here, as in string-theoretic de Sitter,  we find that the window of times with controlled quintessence in our solutions is finite (but large).   

The low energy effective action in perturbative corners of the theory contains canonically normalized scalar fields $\Phi_{\text{c}}$ subject to exponential potential terms of the form 
\beq\label{Vsingle}
V=V_0 e^{\beta_{\text{c}}\Phi_{\text{c}}}
\eeq
A single canonically normalized scalar field with this potential in $d$ spacetime dimensions generates an FRW solution
\beq\label{FRW}
ds^2=-dt^2+a(t)^2 d\vec x^2
\eeq
with a power-law scale factor 
\beq\label{aK}
a(t)= \left(\frac{t}{t_0}\right)^K, ~~~~~ K=\frac{4}{(d-2)\beta_{\text{c}}^2}
\eeq
Accelerated expansion occurs for $K>1$, equivalently $\beta_{\text{c}}^2<4/(d-2)$.  This does not occur for the individual scalar fields corresponding to the dilaton or volume modulus in known weak coupling and large radius limits of string theory.  

Moreover, more general no go theorems for slow roll inflation along certain single-field directions have been proved in various works such as \cite{nogo}, for particular classes of stress energy sources.   
These no go theorems are very useful, but of limited applicability for several reasons.  First, they explicitly restrict the stress energy sources to a subset of those arising in weakly coupled string theory.\footnote{For an early example of slow-roll inflation in brane models see e.g. \cite{Dvali:1998pa}. } More interestingly,  many inflationary mechanisms do not require the individual single-field slow roll conditions to be satisfied.  These include Assisted Inflation \cite{assisted}\ and related versions of multifield inflation \cite{Nflation,Tolley:2007nq}, DBI inflation \cite{DBI}, Locked Inflation \cite{locked}, Thermal Inflation \cite{thermal}, Trapped Inflation \cite{trapped}, Spiral Inflation \cite{spiral}, and interesting combinations such as Unwinding Inflation \cite{unwinding}.  We will exploit some of these possibilities in order to reduce the list of stress-energy sources required to obtain accelerated expansion.  Finally, one may obtain explicit perturbative solutions with strong warping \cite{Douglaswarped}\ in which the scalar sourced by a steep potential of the form (\ref{Vsingle}) varies in an internal spatial direction rather than rolling quickly in time \cite{rollingdS}; this mechanism is also not covered by the existing no go theorems.

In the simplest version of Assisted Inflation \cite{assisted}, for example, one simply considers a set of $N$ fields $\Phi_i$, coupling only through gravity, each with a potential of the form (\ref{Vsingle}).  By rescaling the fields it is straightforward to show that one obtains power-law inflation, but now with
\beq\label{Kmulti}
K_{\text{assisted}} = \sum_i \frac{4}{(d-2)\beta_i^2}
\eeq
So even if none of the fields would support slow roll inflation individually, that is even if $\beta^2_i>4/(d-2)$, the system can still inflate with a sufficiently large number of fields.  

We will find that a generalization of this structure arises in string theory, obtaining more general scaling solutions which we can analyze as in \cite{Tolley:2007nq}.  
In the first set of solutions obtained in \S\ref{SC}, we require three basic ingredients sourcing two fields -- the dilaton and volume modulus --
to obtain accelerating solutions via assisted inflation.  This set of models realizes a wide range of values of $K$, related to the equation of state, the ratio $w$ of energy density to pressure, via $w=-1+\frac{2}{3K}$.  This sequence of possible equations of state arise by varying an integer quantum number in the theory. We first illustrate this with a class of models that accumulate near $w=-1/3$ from below. Next, we construct a sequence that accumulates near $w=-1$; these are potentially interesting for phenomenological applications, providing a new way of realizing dark energy in string theory. These models are somewhat similar to the earliest models of de Sitter spacetime in string theory \cite{SCdS}, but differ in important ways: in addition to realizing a wider variety of equations of state, they are large-radius solutions and more easily controlled including the effects of the large number of species as we discuss below.  In particular, this class of solutions involves the axion fields the dominate the string spectrum in an interesting way, combining assisted inflation, monodromy inflation, and N-flation.  The second set of solutions described in \S\ref{sec:density}\ uses a finite density -- a domain wall network -- to drive accelerated expansion with $1<K<2$. Finally, in \S \ref{sec:applications} we discuss applications of our results as well as other directions for future work. 

\section{Sequences of accelerating FRW solutions}\label{SC}
   
Consider string theory in $D$ dimensions with additional matter sources (which will be orientifolds and fluxes in our examples).  The effective action governing the dilaton and metric in appropriate solutions is 
\be
S =\frac{1}{2\alpha'^{\frac{D-2}{2}}}\int  d^Dx\sqrt{g} \,e^{-2 \phi_\text{s}} \left(\mR-\frac{2}{3}\frac{c-c_{\text{crit}}}{\alpha'}+ 4 (\partial \phi_\text{s})^2 \right) + S_{\text{matter}}\,.
\ee
Here $c$ is the worldsheet matter central charge; this is equal to $(3/2)D$ in the simplest worldsheet-supersymmetric version of supercritical string theory which is free of destabilizing spacetime tachyons \cite{pseudo}.  The string coupling is 
\beq\label{gstring}
g_{\text{s}}=e^{\phi_\text{s}}.
\eeq

In this section, we present a class of examples of accelerated expansion which arise immediately in supercritical limits of string theory. Supercritical string limits are connected to the much more widely studied supersymmetric limits of string theory via various transitions \cite{DimMutation,simeon}.  They appear to be the most generic weakly coupled regions of the full theory, simply because other solutions require turning off the parameter $D-D_\text{crit}$ and negative curvature of the target space.  The vast majority of compact manifolds are negatively curved, and even in $D=10$ these have a supercritical effective central charge $c_{\text{eff}}>10$ precisely computed in \cite{DimMutation}.  At a more basic level, $D=10$ is also a very special choice which at low energy amounts to turning off the leading term in the scalar potential.  Low energy supersymmetry, if observed, would require these special choices.  However, as of this writing no such extension of the Standard Model has been detected, although it remains a viable possibility motivated by various indirect hints from the bottom up.  In any case, our motivation is as much conceptual as phenomenological, and we believe it is highly motivated to analyze the structure of cosmological solutions in string theory using the most tractable tools, including those employed in the present work.        

There are various microscopic (worldsheet) conditions and choices required to formulate this class of theories.   One may consider a variety of consistent GSO projections in this theory, as discussed in e.g. \cite{earlySC, simeon,pseudo}.  The simplest is to make a supercritical analogue of the type IIA or type IIB GSO projections, choosing $D=10$ (mod 16)\footnote{This condition is stronger than $D=2$ (mod 8) because of spin-statistics.  We thank E. Witten for mentioning this to us.} and either a left-right asymmetric or left-right symmetric projection on the worldsheet fermions.  This produces a tachyon-free, modular invariant spectrum of single string states.  We will also require consistency conditions between orientifolds and fluxes of various dimensions, ensuring that the orientifolds not project out the RR fluxes we introduce.       
On a toroidal compactification, this condition is very simple to state:
\be\label{OFcond}
n_{\rm shared} + n_{\rm unwrapped} = -1 \text{ (mod 4)},
\ee
where ``shared" means wrapped by the O-plane and also threaded by the flux; ``unwrapped'' means not wrapped by the O-plane and also not threaded by the flux.  This condition is manifestly invariant under T-duality, and reduces to the standard condition in $D=10$.  On more generic manifolds, the fluxes that would be projected out by this condition can be included, but must live on odd cycles under the orientifold action, and fluxes that are invariant according to this criterion must thread even cycles under the orientifold action.  Another basic microscopic condition is Gauss' law:  the orientifold charge must cancel with anti-orientifolds, branes, or appropriate combinations of fluxes.\footnote{The latter involves NS-NS flux, which is an interesting generalization of \cite{GKP}\ but along with the RR flux it introduces axion couplings which lead to richer dynamics in a different class of models; this case is not included in the present work.} 
In cases with multiple sectors of orientifolds, it is convenient to arrange them to intersect on codimension 0 mod 4, to avoid twisted sector tachyons arising in the sector coming from the product of two orientifold projections.\footnote{This is not a strict consistency condition:  with a more elaborate construction, such sectors may sometimes be projected out.}  

As we will review, the simplest solution to the equations of motion derived from this effective action reproduces the behavior of the string coupling and metric in the standard linear dilaton background \cite{SClindil,joebook}, which has an exact worldsheet description and a coupling that becomes arbitrarily weak at late times.  With two additional sources of stress energy, we will obtain solutions with a controlled $\alpha'$ and loop expansion for a much wider range of equations of state, with accelerated expansion, as we will explain in detail below.        

Let us discuss compactification of the theory.  As a brief warmup, consider a rectangular $n$-dimensional torus, for which we find the following structure.  Let us analyze the $d=(D-n)$--dimensional effective theory in terms of the Einstein frame metric $\hat g_{\mu\nu}$, related to the $D$-dimensional string frame metric via
\be\label{metrics}
ds_{D,\text{str}}^2 = e^{2 \t D} \hat g_{\mu\nu} dx^\mu dx^\nu + \sum_{i=1}^n e^{2\phi_i} dy_i^2
\ee
where $L_i = e^{\phi_i}$ are the sizes of the $n$ circles in string units, and the factor relating the string and Einstein frame metrics   is related to the low energy $d$-dimensional effective coupling:
\be
e^{(d-2) \t D} = e^{2\phi_\text{s} - \sum_i \phi_i}\,.
\ee
We will denote the overall size modulus by 
\be
L^n \equiv e^{\sum_i \phi_i}\,.
\ee

The kinetic terms of the lower dimensional theory are naturally diagonalized in terms of $\t D$ and the $\phi_i$:
\be
S_{\text{eff}}=\frac{1}{2}M_d^{d-2} \int d^d x \sqrt{\hat g} \left(\hat \mR - \sum_{i=1}^n (\hat \partial \phi_i)^2 - (d-2) (\hat \partial \t D)^2 - V_{\text{eff}} \right)
\ee
where
\be\label{Vscmatter}
V_{\text{eff}}= V_{\text{sc}} \,e^{2\t D} + V_{\text{matter}}\,,
\ee
$V_{\text{sc}} \propto (D- D_{\text{crit}})/\alpha'$, $M_d$ is the $d$-dimensional Planck mass, and quantities with `hats' refer to the $d$ dimensional Einstein frame metric $\hat g_{\mu\nu}$.    

More generally, we can compactify on a manifold of volume proportional to $e^{n\phi}$, with $D$-dimensional string frame metric of the form
\be\label{metrics2}
ds_{D,\text{str}}^2 = e^{2 \t D} \hat g_{\mu\nu} dx^\mu dx^\nu + e^{2\phi} \,\gamma_{ij}dy^i dy^j.
\ee
with 
\be
e^{(d-2) \t D} = e^{2\phi_\text{s} - n\phi}\,.
\ee
For these more general spaces,  the effective action for the dilaton and overall volume remains of the form
\be
S_{\text{eff}}=\frac{1}{2} M_d^{d-2} \int d^d x \sqrt{\hat g}\left(\hat \mR - n (\hat \partial \phi)^2 - (d-2) (\hat \partial \t D)^2 - V_{\text{eff}} \right),
\ee
Additional dependence on other fields will generically arise, depending on the sources and initial conditions.
However for the sake of simplicity, we will consider simple setups where anisotropies do not participate in the dynamics.  

\subsection{Scaling solutions}\label{subsec:scaling}

Before beginning our analysis of explicit microscopic models, let us discuss some general properties of the FRW cosmologies that we will find. We focus on isotropic models, described by the two fields $\t D$ and $\phi$ above. 
As discussed above, the potential for $\phi$ and $\t D$ descending from the $D$ dimensional theory is a sum of exponentials,\footnote{It is interesting to include other types of fields such as axions, with power law and sinusoidal contributions to $V$, but they will not play a role in the present work.}
\be\label{Veff}
V_{\text{eff}}=\sum_i \,V_i \,e^{\alpha_i \t D + \beta_i \phi}\,.
\ee
The exponents $(\alpha_i, \, \beta_i)$, which for now we will take to be arbitrary, are determined by the choice of sources. The supercritical potential reviewed above has $(\alpha,\,\beta)=(2,0)$.

Other contributions to the effective potential arise from various sources.  In the absence of strong warping \cite{Douglaswarped}, a simplification we will ensure using the methods developed in \cite{micromanaging},  these are straightforward to derive from the higher dimensional theory.\footnote{For a pedagogical review of these and other terms in the string theory effective potential see \cite{TASIlectures}.}  
Internal curvature gives $(\alpha,\,\beta)=(2,-2)$, a D-brane/orientifold wrapping $n_{\text{B}}$ internal dimensions has $(\alpha,\,\beta)=(\frac{d+2}{2},n_{\text{B}}-\frac{n}{2})$ and a p-form RR flux contributes $(\alpha,\,\beta)=(d,n-2p)$. 

The simplest solutions that we will obtain are scaling solutions where $a(t) \sim t^K$ and the scalar fields depend logarithmically on time.
To understand how they arise, consider a potential with just two terms.
We have a scaling transformation
\be
\t D \,\to\,\t D  -2\frac{\beta_1-\beta_2}{\alpha_2 \beta_1-\alpha_1 \beta_2}\, \lambda\;,\;\phi\,\to\,\phi + 2\frac{\alpha_1 - \alpha_2}{\alpha_2 \beta_1-\alpha_1 \beta_2}\,\lambda
\ee
under which the action transforms homogeneously (with an appropriate transformation of the metric) \cite{Tolley:2007nq}.
This extends to additional terms in the potential if the additional exponents satisfy
\be\label{eq:betai}
 (\beta_1 - \beta_2)\alpha_i-(\alpha_1 - \alpha_2) \beta_i = \alpha_2 \beta_1-\alpha_1 \beta_2\,,
\ee
for $i \ge 3$.
Assuming that this relation is satisfied, we look for a scaling solution in which each term in the action evolves like $(t_0/t)^2$.  In terms of the FRW scale factor and the scalar fields, this translates into the ansatz
\be
a(t) =\left(\frac{ t}{t_0}\right)^K\;,\; \t D(t) = D_0  -2\frac{\beta_1-\beta_2}{\alpha_2 \beta_1-\alpha_1 \beta_2} \,\log\,\frac{ t}{t_0}\;,\;\phi(t) = \phi_0 + 2\frac{\alpha_1 - \alpha_2}{\alpha_2 \beta_1-\alpha_1 \beta_2} \,\log\,\frac{ t}{t_0}\,,
\ee
which we plug into the equations of motion
\bea\label{eom}
&& (d-1) (d-2) \left(\frac{\dot a}{a} \right)^2=(d-2) \dot{\t D}^2 + n \dot \phi^2 +2 \,V_{\text{eff}} \nonumber\\
&& (d-2) \left(\ddot{\t D}+(d-1) \frac{\dot a}{a} \dot{\t D} \right) + \partial_{\t D} V_{\text{eff}}=0 \\
&& n \left(\ddot{\phi}+(d-1) \frac{\dot a}{a} \dot{\phi} \right)+ \partial_{\phi} V_{\text{eff}}=0 \nonumber\,.
\eea

It is useful to proceed analytically and redefine fields to pick out a field $\Phi$ along the scaling direction, and a transverse field $\sigma$, as in \cite{Tolley:2007nq}. 
The scaling direction $\Phi$ and the orthogonal combination $\sigma$ are given by
\be\label{eq:Phisigma-def}
\Phi = (d-2) (\beta_1 - \beta_2) \t D - n (\alpha_1- \alpha_2) \phi\;,\;\sigma = (\alpha_1 - \alpha_2) \t D + (\beta_1 - \beta_2) \phi\,.
\ee
This transformation is such that the effective potential depends on the scaling direction $\Phi$ only through an overall factor,
\be\label{eq:generalVeff}
V_{\text{eff}}=e^{\beta_\Phi\, \Phi}\,e^{-\gamma \sigma}\left( V_1 + V_2 \,e^{-\sigma}+\sum_{i\ge 3}\,V_i\,e^{- \frac{\alpha_1 - \alpha_i}{\alpha_1 - \alpha_2} \sigma} \right)
\ee
and the kinetic terms are diagonal,
\be\label{eq:twofieldL}
(d-2) (\hat \partial \t D)^2 + n (\hat \partial \phi)^2 =\Delta^{-1} \left(n(d-2) \, (\hat \partial \sigma)^2+(\hat \partial \Phi)^2 \right)\,.
\ee
We have defined
$\Delta \equiv n (\alpha_1- \alpha_2)^2+(d-2)(\beta_1 - \beta_2)^2$
and the exponents
\bea\label{eq:exponents}
\beta_\Phi&=&\Delta^{-1}\left(\alpha_2 \beta_1 - \alpha_1 \beta_2\right)\nonumber\\
\gamma& = &\Delta^{-1}\left(n \alpha_1 (\alpha_2 - \alpha_1)+(d-2) \beta_1 (\beta_2 - \beta_1)\right)\,.
\eea

In this form it is easy to find the cosmological solution. Changing to the canonically normalized $\Phi_{\text{c}}=\Phi/\sqrt{\Delta}$ and defining $\beta_{\Phi_\text{c}}= \sqrt{\Delta} \beta_\Phi$,  we obtain a scaling exponent for $a(t)=(t/t_0)^K$ (\ref{aK}),
\be\label{eq:Kgeneral}
K = \frac{4}{(d-2) \,\beta_{\Phi_\text{c}}^2}= \frac{4}{d-2}\,\frac{\Delta}{(\alpha_1 \beta_2 - \alpha_2 \beta_1)^2}
\ee
and the scaling direction evolves according to
\be
\Phi_\text{c}= \Phi_{\text{c},0} - \frac{2}{\beta_{\Phi_\text{c}}}\,\log\,\frac{t}{t_0}\,.
\ee
Stability of the $\sigma$ direction requires that 
\be\label{eq:tV}
\t V \equiv e^{-\gamma \sigma}\left( V_1 + V_2 \,e^{-\sigma}+\sum_{i\ge 3}\,V_i\,e^{- \frac{\alpha_1 - \alpha_i}{\alpha_1 - \alpha_2} \sigma} \right)
\ee
admits a minimum, and the value at the minimum, $V_*$ has to be positive. 
It is related to the coefficient $\Phi_{\text{c},0}$ by
\be\label{Vstar}
e^{ \beta_{\Phi_\text{c}} \Phi_{\text{c},0}} (V_* t_0^2)=\frac{1}{2}(d-2) K \left((d-1)K-1 \right) \,.
\ee

In the string-theoretic models that we will present shortly, the first term $V_1=D-D_{\text{crit}}$ comes from the supercritical potential (\ref{Vscmatter}), the intermediate term $V_2 = -|V_\text{O}|$ is negative and proportional to the orientifold tension, and the third contribution $V_3= Q^2$ comes from fluxes, with $\frac{\alpha_1 - \alpha_i}{\alpha_1 - \alpha_2} =2$.
This potential admits a minimum for $\sigma$ with positive energy as long as
\be\label{eq:window1}
1< \frac{4(D-D_\text{crit}) Q^2}{V_\text{O}^2}< \frac{(1+\gamma)^2}{\gamma(2+\gamma)}\,.
\ee
and the value at the minimum
\be\label{eq:sigmamin}
e^{-\sigma_*}= \frac{1}{2(2+\gamma)} \,\frac{|V_\text{O}|}{Q^2} \left(1+ \gamma + \sqrt{1- \gamma(2+\gamma) \left(4 \frac{V_\text{sc} Q^2}{V_\text{O}^2}-1\right)} \,\right)\,.
\ee
It is also useful to note that in models where the supercritical potential participates in the scaling solution (as will be the case in the constructions below), the lower dimensional effective coupling has a simple power-law behavior,
\be\label{eq:scaling-geff}
g_\text{eff}^2=\frac{g_\text{s}^2}{L^n}=g_{\text{eff},0}^2 \left( \frac{t_0}{t}\right)^{d-2}\,.
\ee
This follows from the scaling of the first term in the potential, $e^{2 \t D}\sim 1/t^2$, and the definition of $\t D$.

Finally, note that if $\alpha_2 \beta_1 - \alpha_1 \beta_2 \to 0$ with $\Delta$ finite, $K$ diverges. This corresponds to a de Sitter solution. The simplest way to see this is to go back to (\ref{eq:generalVeff}) and note that $\beta_\Phi \to 0$ in this limit. Therefore, $\Phi$ becomes a flat direction, and the stabilization of $\sigma$ gives rise to de Sitter with cosmological constant proportional to $V_*$.

In what follows we will construct string theory examples that lead to stable accelerating cosmologies of the form we have just derived, with specific results for $K$ (and hence the equation of state) as a function of integer parameters.

\subsection{A warm-up toy model (toroidal compactifictions)}\label{subsec:compactifiction}

We will present our main sequence of models in \S\S \ref{subsec:simpletori}, \ref{subsec:tori}.  These will satisfy the requisite microscopic consistency conditions described around (\ref{OFcond}).  Before turning to these complete models, we will develop some intuition in this section by working out a sequence of supercritical string-inspired bottom-up models which have some of the essential features of our ultimate examples, but do not in themselves respect some of the microscopic consistency conditions.  As will become clear, the basic mechanism at play in these simple toy models will be used in \S\S \ref{subsec:simpletori} and \ref{subsec:tori}, where we build fully top-down models satisfying all consistency conditions, with similar results.  

First, we notice that for $V_{\rm matter}=0$ in our potential (\ref{Vscmatter}), the radii $\phi_i$ are not sourced and the theory admits an FRW solution expanding linearly with time:
\be
a(t) =\frac{t}{t_0} \,,\quad
\t D=\log \frac{d-2}{\sqrt{2 V_\text{sc} t_0^2}}- \log\frac{t}{t_0}\,.
\ee
That is, the theory without additional sources already admits a $K=1$ solution, driven by the potential and kinetic energy of the $d$-dimensional effective string coupling.  Given that, it is natural to check whether including the potential $V_{\rm matter}(\phi_\text{s},\phi_i)$ can assist in producing accelerated expansion ($K>1$) along the lines of \cite{assisted}.

One of the simplest ways to achieve this in a toroidal compactification is to introduce orientifold planes wrapping $n_\text{O}$ directions out of the $n$ internal dimensions, and to include $Q_p$ units of magnetic RR fluxes threading $p$-dimensional cycles.  For simplicity let us consider the isotropic case, in which we arrange the O-planes and fluxes symmetrically across all $n$ internal dimensions.  Since we are looking for the simplest toy model, we will for the moment ignore the consistency conditions among the GSO projection, the O-planes, and the fluxes.  We will adjust the model to account for these conditions including all the effects of the full orientifold group in the complete examples below; in the full models we will also address angular moduli that are ignored in the present section.    

The potential $V_\eff$ is then of the form \eqref{Veff}, with three terms given by the supercritical potential, O-planes, and fluxes.  As mentioned before, their $(\alpha,\beta)$ exponents are
\be
\alpha_1=2\;,\;\beta_1=0\;,\;\alpha_2 = \frac{d+2}{2}\;,\;\beta_2=n_\text{O}- \frac{n}{2}\;,\;\alpha_3=d\;,\;\beta_3=n-2p\,.
\ee
The condition \eqref{eq:betai} for a scaling solution becomes
\be\label{sccond}
n=p+n_\text{O} \,.
\ee
We note that this condition cannot be satisfied directly in the microscopic theory for a very basic reason -- the consistency condition \eqref{OFcond} implies that $n-(p+n_\text{O})$ is odd.  In \S \ref{subsec:microconsistency} we will introduce a microscopically consistent ``flux averaging'' technique to obtain an effective flux quantum number $p_\eff$ that satisfies \eqref{sccond}.  For now, let us proceed with our toy model, since its behavior as a function of $p$ will be similar to that of complete models as a function of $p_{\text{eff}}$.  

As shown generally in \S\ref{subsec:scaling}, it is useful to rewrite the potential in terms of the fields $\Phi$ (which rolls down the potential) and a transverse field $\sigma$ (which is static) as in \eqref{eq:generalVeff}:
\be\label{Vefftoy}
V_\eff= e^{(2p-n)\Delta^{-1} \Phi}\,e^{-n(d-2)\Delta^{-1}\sigma}\, \left(V_\text{sc} - |V_\text{O}| e^{-\sigma} +  Q_p^2 e^{-2 \sigma}\right) \,,
\ee
where as before $\Delta$ is given by
\be
\Delta = \frac{1}{4} (d-2) \left((d-2) n+(n-2 p)^2\right) \,.
\ee
Therefore, the $\sigma$ field transverse to the rolling $\Phi$ direction may be stabilized by the three-term structure in \eqref{Vefftoy} for coefficients in the window (\ref{eq:window1}). For a large number of internal dimensions, the orientifold tension grows as $|V_\text{O}| \sim 2^{n/4}$ \cite{SCdS}, and then (\ref{eq:window1}) is satisfied for fluxes $Q \sim |V_\text{O}| \sim 2^{n/4}$ (up to factors that do not grow exponentially with $n$).

The model generates power-law FRW expansion $a(t)=(t/t_0)^K$ with $K$ given by \eqref{eq:Kgeneral}.  Plugging in the specific values, we get
\be\label{Kns2}
K=1+\frac{n (d-2)}{(n-2p)^2}.
\ee
As $p$ approaches $n/2$, the solution becomes rapidly accelerating and approaches a de Sitter cosmology.
In our complete sequence of solutions below, for which we will establish microscopic consistency and control, we will find a similar formula exhibiting a discrete sequence of equations of state.

\subsection{Microscopic consistency and flux averaging}\label{subsec:microconsistency}

As we saw above, the toy model we just developed fails to be a complete solution because the condition for a scaling solution \eqref{sccond} cannot be satisfied directly in the microscopic theory; the flux quantum number $p$ would need to be odd in type IIA and even in type IIB.  The theory provides an elegant way around this difficulty, however, which we will explain next.  We will refer to this as ``flux averaging''.

Let us first consider a setup which has two RR fluxes, one threading a $p_1$-dimensional cycle and another threading an orthogonal $p_2$-dimensional cycle.  This potentially introduces an anisotropy between the $p_1$ and $p_2$ directions, so we will include separate size moduli $L_1$ and $L_2$ for them.  The anisotropy between the $p_1+p_2$ directions and the rest on the internal manifold can be avoided by e.g.\ symmetrically arranging the $p_1+p_2$ directions across all internal dimensions.  We will do this in the detailed top-down models in the next subsections.

With these specifications, the flux potential is of the form
\be
V_{\rm flux} = \(\frac{g_{\text{s}}^2}{L^n}\)^{d/(d-2)} L^n \(\frac{Q_1^2}{L_1^{2 p_1}} +\frac{Q_2^2}{L_2^{2 p_2}}\) \,,
\ee
where $Q_1$, $Q_2$ are the flux numbers, and the overall size modulus $L$ is defined as $L^{p_1+p_2} = L_1^{p_1} L_2^{p_2}$.  Using this we may rewrite the flux potential as
\be
V_{\rm flux} = \(\frac{g_\text{s}^2}{L^n}\)^{d/(d-2)} L^n \(\frac{Q_1^2}{L^{p_1+p_2}} \frac{L_2^{p_2}}{L_1^{p_1}} +\frac{Q_2^2}{L^{p_1+p_2}} \frac{L_1^{p_1}}{L_2^{p_2}}\) \,.
\ee
We see that the combination $L_1^{p_1}/L_2^{p_2}$ is stabilized at order $Q_1/Q_2$, and the flux potential becomes
\be
V_{\rm flux} = \(\frac{g_\text{s}^2}{L^n}\)^{d/(d-2)} L^n \frac{2 Q_1 Q_2}{L^{p_1+p_2}} \,.
\ee
Therefore, after stabilizing the anisotropic direction the two fluxes contribute to the potential as if they were both $p_\eff$-form fluxes with $p_\eff=(p_1+p_2)/2$, effectively averaging the flux.  

It is now clear that we can get $p_\eff$ even if the microscopic theory does not allow a fundamental RR field strength of this rank.  For example, using a pair of $p$-form and $(p+2)$-form fluxes (if they are compatible with other sources such as the orientifolds) in this way, we may get $(p+1)$-form fluxes.  In type IIA, we can get odd $p_\eff$ starting from even $p$, and viceversa for type IIB.

We will actually use a slightly generalized version of this flux averaging. Let us consider two RR fluxes threading a total of $n_1+n_2$ directions.  The first flux wraps $p_{11}$ of the $n_1$ directions and $p_{12}$ of the $n_2$ directions.  Similarly, the second flux wraps $p_{21}$ of the $n_1$ directions and $p_{22}$ of the $n_2$ directions.  We arrange the $p_{11}$ directions symmetrically out of the $n_1$ directions, and do the same for $p_{12}$, $p_{21}$, and $p_{22}$.

The flux potential becomes
\be
V_{\rm flux} = \(\frac{g_\text{s}^2}{L^n}\)^{d/(d-2)} L^n \(\frac{Q_1^2}{L_1^{2p_{11}} L_2^{2p_{12}}} +\frac{Q_2^2}{L_1^{2p_{21}} L_2^{2p_{22}}}\) \,,
\ee
where $L^{n_1+n_2} = L_1^{n_1} L_2^{n_2}$.  Again, this potential stabilizes the combination $L_1^{p_{11}-p_{21}} L_2^{p_{12}-p_{22}}$ at order $Q_1/Q_2$ and gives an effective rank
\be\label{peff}
p_\eff = (n_1+n_2) \frac{p_{11} p_{22} - p_{12} p_{21}}{n_1 (p_{22}-p_{12}) +n_2 (p_{11}-p_{21})} \,.
\ee
As a trivial check, this reduces to $(p_1+p_2)/2$ for $n_1=p_{11}=p_1$, $n_2=p_{22}=p_2$, and $p_{12}=p_{21}=0$.


\subsection{Toroidal orientifold models generating acceleration with $w\lesssim -1/3$}\label{subsec:simpletori}

In this section we will construct a family of supercritical models on $\mathbf{R}^{d-1,1} \times \mathbf{T}^n$ which are microscopically complete and similar to the toy models above for the case that the orientifold planes wrap a large fraction of the torus ($n_\text{O}$ of order $n$ in the notation of \S\ref{subsec:compactifiction}).  This yields a sequence of possible equations of state accumulating near $w\to -1/3$.  In the next section we will produce a somewhat more involved sequence of models with $w$ accumulating near $-1$, similar to the $n_\text{O}\approx n/2$ toy examples.    

We will first describe a class of models in some generality, and then make specific choices which satisfy all the required microscopic consistency conditions.  
Let us divide the $n$ toroidal directions equally into $m$ ``blocks'',  each block consisting of $n'\equiv n/m$ directions.  Let there be orientifold planes wrapping $m_\text{O}$ blocks, chosen symmetrically from the total $m$ blocks, along with anti-orientifolds separated from them in the internal dimensions.  These are $O(d-1+n_\text{O})$ planes with $n_\text{O}=m_\text{O} n'$, introduced by modding out the worldsheet theory by an action of $\Omega I_{n'(m-m_\text{O})} $
times and appropriate power of $(-1)^{F_L}$ which flips the sign of spacetime spinors from worldsheet left-movers.  Here $\Omega$ is worldsheet parity and $I_j$ indicates reflection on $j$ coordinates transverse to the O-plane.  
The simplest consistent models have O-planes which intersect on codimension $0$ (mod 4) which means $n'$ is even.  Therefore $n_\text{O}$ is even and we are in type IIA for odd $d$ and in type IIB for even $d$.

We must consider all elements of the orientifold group that we have prescribed.  Since we have distributed the orientifolds symmetrically among the blocks, this includes orbifold elements $I_{2n'}$ which invert $2n'$ directions, and their products.  In order to avoid tachyons in the twisted sectors of this orbifold group we require $n'$ to be even.  The full orientifold group as just specified generically also generates additional sectors of O-planes of higher and lower dimensionality than the original set of  $O(d-1+n_\text{O})$ planes (generated by actions of $I_{2n'}$ elements transverse to or parallel to a given O-plane).  We will be interested in a controlled large-radius regime, so the O-planes of smaller dimension than $d-1+n_\text{O}$ will automatically be negligible in the dynamics.  The higher-dimensional O-planes would dominate over lower dimensional ones at large volume, for models in which they are generated.  For the cases when this occurs we can neutralize this contribution in two ways. One is to include D-branes which cancel their tension, leading to a stable configuration.\footnote{It is also interesting to note that in high transverse dimensionality (as we will have in our blocks), the classical forces between masses and charges are suppressed \cite{EmparanD}}.  This mechanism does not apply to a completely spacefilling orientifold generated by the action $\Omega$ with no reflection.  To address that if it is generated by the full orientifold group, we can include a shift halfway around a direction of the torus with the element $\Omega$ and with each element of the form $I_{2n'}$ which acts on that direction.  This removes the spacefilling O-plane, since this element of the orientifold group acts freely, but preserves the fixed points introduced by the original O-planes specified above.      

With this distribution of orientifolds, let us next check that the angular moduli of the torus do not become unstable.  With orientifolds, which carry negative tension, wrapping sub-tori of the $T^n$ , it is energetically favorable for angular moduli to turn on in such a way as to increase the volume wrapped by the O-planes at fixed volume of the $T^n$.  However, such angular moduli -- which are off-diagonal terms in the metric of the form $dx \,dy$, where $x$ and $y$ are from different blocks -- are automatically projected out in our model.   The orientifold group includes $\mathbf{Z}_2$ orbifold elements which acts as a reflection on all coordinates in any two pairs of blocks.  Each angular modulus of the form just discussed is projected out by this group. 

To introduce RR fluxes, let us further divide each block into $n_1$ and $n_2$ directions, $n'=n_1+n_2$.  We will denote the radii of these two directions in string units as $L_1=e^{\phi_1}, L_2=e^{\phi_2}$ as above in \S\ref{subsec:microconsistency}.  Two kinds of RR fluxes are turned on; the first threads $p_{11}$ of the $n_1$ directions and $p_{12}$ of the $n_2$ directions in each block, so it is a $p_1=m(p_{11}+p_{12})$ form flux.  The second flux is a $p_2=m(p_{21}+p_{22})$ form wrapping $p_{21}$ of the $n_1$ directions and $p_{22}$ of the $n_2$ directions.  We arrange the $p_{ij}$ directions symmetrically out of the $n_j$ directions.

As derived in \eqref{peff}, we may balance the potential terms from the two fluxes and obtain an effective $p_\eff$-form where
\be\label{peffm}
p_\eff = m n' \frac{p_{11} p_{22} - p_{12} p_{21}}{n_1 (p_{22}-p_{12}) +n_2 (p_{11}-p_{21})} \,.
\ee
In order to find a power-law solution, we need to satisfy the scaling condition \eqref{sccond}, $n = p_\eff + n_\text{O}$.
Having done this, we will find results similar to those of the simpler toy model of \S\ref{subsec:compactifiction}.

We now present a set of solutions along these lines which satisfies our microscopic conditions.
We choose
\be\label{p1ink}
n_1=n_2=3k\;,\;p_{11}=4\;,\;p_{21}=0\;,\;p_{12}=1\;,\;p_{22}=9\,.
\ee
which gives $p_\eff=\frac{m}{k} n'$, with $n'=n_1+n_2=6k$. Setting for simplicity $m=k$, obtains $n=mn'=6k^2 $, $p_\eff=6k$ and $n_\text{O}=n-p_\eff=(m-1)n'=6k(k-1)$.
The number of blocks wrapped by the O-planes is $m_\text{O}=m-1$.  This means that they cannot generate higher-dimensional O-planes, and will only generate lower-dimensional O-planes (differing from these original ones by even numbers of blocks) which contribute subdominantly at large radii.

Let us check the remaining microscopic consistency conditions. 
We must make sure that the orientifold actions do not project out the fluxes we use.  For either flux the condition \eqref{OFcond}\ becomes
\be
k=-1 \text{ (mod 4)} \,.
\ee
And finally we need $D=d+n=10$ (mod 16) for our modular invariant GSO projection, which is satisfied in $d=4$ by any $k=-1$ (mod 4).

Solving for the classical dynamics of this model following the method described above in \S\ref{subsec:scaling}, we obtain a perturbatively stable solution exhibiting accelerated expansion $a(t) = (t/t_0)^K$ with  
\be\label{Ksimple}
 K= 1+\frac{10}{27 (k-2)^2} \,.
\ee
In the limit of large $k$, this sequence of models approaches $K=1$, which corresponds to $w=-1/3$.

Let us explain how this comes about in some detail.  
The kinetic terms are as in \eqref{eq:twofieldL}, and the potential is given by
\be
V_\eff = e^{\Phi} e^{-\gamma\sigma_1} \[(D-D_\text{crit}) -|V_\text{O}| e^{-\sigma_1-\sigma_2} + e^{-2\sigma_1}(Q_1^2 e^{-\gamma_1 \sigma_2} + Q_2^2 e^{-\gamma_2 \sigma_2})\] \,,
\ee
with the relation between the dilaton and radii 
\bea\label{eq:relation1}
\phi_\text{s} &=& -\frac{1}{k-2} \Phi - \(1+\frac{54 (k-2)}{27 k^2-108 k+118}\) \sigma_1 - \frac{k}{k-2} \sigma_2 \nonumber\\
\phi_1 &=& -\frac{2}{9k (k-2)} \Phi - \frac{12 (k-2)}{k (27 k^2-108 k+118)} \sigma_1 + \frac{2}{3k (k-2)} \sigma_2 \\
\phi_2&=& -\frac{1}{9k (k-2)} \Phi -\frac{6 (k-2)}{k (27 k^2-108 k+118)} \sigma_1 -\frac{4}{3k (k-2)} \sigma_2 \nonumber\,,
\eea
and the exponents 
\be
\gamma= \frac{20}{27 k^2-108 k+118}\;,\;
\gamma_1= 2+\frac{20}{3(k-2)}\;,\;
\gamma_2= 2-\frac{20}{k-2} \,.
\ee

We easily find a stable minimum of the potential in the $\sigma_1,\sigma_2$ directions, which can be understood rather simply as follows.  Since $\gamma$ is tiny at large $k$, we can neglect its effects on the stabilized values of the moduli near the minimum that we will find (although it comes into the value of the potential there).  Now consider the potential as a function of the two fields  $\sigma_+ \equiv \sigma_1+\sigma_2$ and $\sigma_2$.  The two flux terms stabilize $\sigma_2$, near zero if we take for simplicity $Q_1^2 \approx 3Q_2^2 \approx Q^2$.  The remaining potential is then of the form (\ref{eq:tV}), and restricting the parameters to the window (\ref{eq:window1}) gives a minimum for $\sigma_+$ at (\ref{eq:sigmamin}).  Explicitly, if we choose the flux numbers $Q \sim \frac{|V_\text{O}|}{\sqrt{D-D_\text{crit}}} \sim \frac{2^{n/4}}{n^{1/2}} $, we obtain a minimum with\footnote{Recall that $V_*$ was introduced below (\ref{eq:tV}).}
\be\label{eq:sigma1Vstar}
e^{\sigma_1} \sim \frac{|V_\text{O}|}{D-D_\text{crit}}\sim \frac{2^{n/4}}{n} \;,\;V_* \sim \epsilon (D-D_\text{crit}) M_4^2\,,
\ee
where $\epsilon$ depends on how close the flux is tuned to the lower bound of the window (\ref{eq:window1}). Furthermore, the combination $e^{\Phi_0}t_0^2$ that appears in $\Phi= \Phi_0 - 2 \log(t/t_0)$ is fixed in terms of $V_*$ by the relation (\ref{Vstar}). 

Replacing these results into (\ref{eq:relation1}), we finally arrive to the solution for the string coupling and radial moduli for this class of models (expanded at large $k$):
\be\label{eq:solmodel1}
g_\text{s} \sim 2^{-\frac{3}{2}k^2} (V_* t^2)^{\frac{1}{k}}\;,\; L_1 \approx L_2^2 \sim 2^{-\frac23} (V_* t^2)^{\frac{2}{9k^2}}\,,\;L \sim 2^{-\frac12}  (V_* t^2)^{\frac{1}{6k^2}}
\ee
and the effective coupling  (recall that $L^n=L_1^{mn_1}L_2^{mn_2}$)
\be\label{eq:geffmodel1}
g_\text{eff}^2=\frac{g_\text{s}^2}{L^n} \sim \frac{1}{V_* t^2}\,.
\ee
The time-dependence of the effective coupling was derived in (\ref{eq:scaling-geff}), and can also be seen directly from the linear combination $2\phi_s-m n_1 \phi_1 - m n_2 \phi_2$ in (\ref{eq:relation1}).\footnote{This arises from subleading $1/k$ contributions not shown in (\ref{eq:solmodel1}).}

The string coupling and radial moduli increase with time, and we will find a window of large radius and weak coupling.
We will explain that in detail below in 
\S\ref{subsec:control}, where we will first lay out the general criteria required to establish control in our large-$D$ regime
before applying them to our specific sequences of models.

\subsection{A sequence  approaching $w=-1$}\label{subsec:tori}

In this section, we will consider another sequence of models which produces stronger accelerated expansion, approaching $w\to -1$ ($K\to\infty$) at large $D$.  These give a microscopically consistent realization of the $p\approx n/2$ regime of the warmup models in \S\ref{subsec:compactifiction}.  

To begin, we will consider a family of supercritical models on $\mathbf{R}^{d-1,1} \times \mathbf{T}^n$, with the same block structure we described at the beginning of \S\ref{subsec:simpletori}.  In these models, we will find that the many Ramond-Ramond fields of the supercritical theory contribute large radiative corrections unless we lift them using additional ingredients (see \S\ref{subsec:control} for a detailed discussion).  We will explain how to do so in a generalization of the model that combines assisted inflation with axion monodromy on a more general internal geometry.  In particular, it will be interesting to make use of the $~2^D$ axion fields which dominate the string spectrum.     

\subsubsection{First attempt}

Consider to begin with a discrete family of these toroidal orientifold models parametrized by an arbitrary even integer $k\ge2$, now with
\ba
n_1&=16k+1 \,,&n_2&=16k-7 \,,\nonumber\\
p_{11}&=16k \,,&p_{12}&=1 \,,\label{p1inkII}\\
p_{21}&=0 \,,&p_{22}&=16k-7 \nonumber\,.
\ea
Then from \eqref{peffm} we find
\be
p_\eff = \frac{m}{64k^2-28k-1} (32k^2-14k)n'
\ee
where $n'=n_1+n_2=32k-6$.  Since the scaling condition $n=p_\eff+n_\text{O}$ demands that $p_\eff$ be an integer, we make the simplest choice
\be
m=64k^2-28k-1
\ee
which leads to 
\ba\label{nink}
n&=mn'=(64k^2-28k-1)(32k-6) \,, \nonumber \\
p_\eff&=(32k^2-14k)(32k-6) \,,\\
n_\text{O}&=n-p_\eff=(32k^2-14k-1)(32k-6)\nonumber \,.
\ea
In particular, this means that the number of blocks wrapped by the O-planes is $m_\text{O}=2k(16k-8)$.

Let us now reiterate and finish checking the various microscopic consistency conditions for these models.  First, for the GSO projection we need $d+n=10$ (mod 16).  From \eqref{nink} and the fact that $k$ is even, we easily find $n=6$ (mod 16), and therefore $d=4$ (mod 16).  We will focus on $d=4$ below for simplicity.
Second, $n'=32k-6$ is even,
which ensures that the $\mathbf{Z}_2$ orbifold elements of the orientifold group do not generate twisted tachyons. Third, we need to satisfy the consistency condition \eqref{OFcond} between the O-planes and RR fluxes.  Consider the $p=m(p_{11}+p_{12})$ form flux with any O-plane.  We have $n_{\rm shared}=m_\text{O}(p_{11}+p_{12})$ and $n_{\rm unwrapped}=(m-m_\text{O})(n'-p_{11}-p_{12})$.  Plugging in the numbers, we find 
$$
n_{\rm shared}+n_{\rm unwrapped}=-p_{11}-p_{12}+(64k^2-28k)(16k-3) \,,
$$
and therefore the condition \eqref{OFcond} becomes $p_{11}+p_{12}=1$ (mod 4) which is certainly satisfied by \eqref{p1ink}.  Similarly, if we consider the other flux we get $p_{21}+p_{22}=1$ (mod 4) which is satisfied by \eqref{p1ink}.

We are now ready to analyze the dynamics of these models.  Since each block is divided into $n_1$ and $n_2$ directions, they can have different size moduli.  Let us denote their lengths by $L_1$ and $L_2$.  These lengths are the same for all blocks, because we have arranged all sources symmetrically.  Therefore, our models have three moduli $g_\text{s}$, $L_1$, and $L_2$.  The potential is of the form (focusing on $d=4$)
\be\label{eq:Veff2}
V_\eff = \(\frac{g_\text{s}^2}{L^n}\)^{2}\((D-D_\text{crit})\frac{L^n}{g_\text{s}^2} -|V_\text{O}| \frac{L_1^{m_\text{O} n_1} L_2^{m_\text{O} n_2}}{g_\text{s}} +\frac{Q_1^2 L^n}{L_1^{2m p_{11}} L_2^{2m p_{12}}} +\frac{Q_2^2 L^n}{L_1^{2m p_{21}} L_2^{2m p_{22}}}\) \,,
\ee
where $L$ is the overall size modulus defined as $L^{n'}=L_1^{n_1} L_2^{n_2}$, and $Q_1$, $Q_2$ are the flux numbers.  We may rewrite this potential in terms of a scaling field $\Phi$ and two transverse fields $\sigma_1$, $\sigma_2$, and arrive at a form similar to \eqref{eq:generalVeff}:
\be\label{eq:Veff3}
V_\eff = e^{\Phi} e^{-\gamma\sigma_1} \[(D-D_\text{crit}) -|V_\text{O}| e^{-\sigma_1-\sigma_2} + e^{-2\sigma_1}(Q_1^2 e^{-\gamma_1 \sigma_2} + Q_2^2 e^{-\gamma_2 \sigma_2})\] \,.
\ee
The new fields and exponents $\gamma$ and $\gamma_i$ are uniquely fixed by comparing (\ref{eq:Veff2}) and (\ref{eq:Veff3}), and requiring that $\Phi$ and $\sigma_i$ have diagonal kinetic terms.
The new and old fields are related by
\bea\label{eq:change-var}
\phi_\text{s} &=&2 k (16 k-7) \Phi + \left(16 k-\frac{16 k-7}{ 16 k^2 (16 k-7)-4 k+1}-8 \right) \sigma_1+ \left(4 k (16 k-7)-1 \right) \sigma_2 \nonumber\\
\phi_1 &=& \frac{2 k-1}{4 (16 k-7) k-1} \,\Phi +\frac{4 k-2}{ 16 k^2 (16 k-7)-4 k+1} \sigma_1 +\frac{4 k}{16 k+1} \sigma_2 \\
\phi_2&=& \frac{2 k}{4 k (16 k-7)-1} \,\Phi +\frac{4 k}{16 k^2 (16 k-7)-4 k+1} \sigma_1 -\frac{4 k-2}{16 k-7} \sigma_2 \nonumber\,,
\eea
and the exponents $\gamma$, $\gamma_1$, and $\gamma_2$ are
\ba\label{eq:exponentsk}
\gamma&= \frac{512 k^3-352 k^2+48 k+4}{256 k^3-112 k^2-4 k+1} \,, \nonumber \\
\gamma_1&= \frac{2048 k^3-1536 k^2+248 k+18}{256 k^2-96 k-7} (64 k^2-28 k-1) \,,\\
\gamma_2&= -(8 k-2) (64 k^2-28 k-1)\nonumber \,.
\ea

This potential stabilizes $\sigma_1$ and $\sigma_2$ and leads to a power-law scaling solution with
\be\label{Ktoroidal}
K= 4 k + \frac{1}{64 k^2-28 k-1} \,.
\ee
For $Q_1 \sim Q_2$, $\sigma_2$ has a stable minimum at the origin. The remaining potential for $\sigma_1$ is then of the form (\ref{eq:tV}) and admits a minimum for a suitable window of the coefficients (\ref{eq:window1}). Taking
$Q \sim \frac{1}{\sqrt{D-D_\text{crit}}}|V_\text{O}| \sim \frac{2^{n/4}}{n^{1/2}}$ obtains
\be\label{eq:sigmamodel2}
e^{\sigma_1} \sim \frac{|V_\text{O}|}{D-D_\text{crit}}\;,\;V_* \sim \epsilon(D-D_\text{crit}) \left(\frac{D-D_\text{crit}}{|V_\text{O}|} \right)^\gamma M_4^2
\ee
with $\gamma \approx 2$ at large $k$ and, as before, $\epsilon$ depends on how close to the lower bound of the window (\ref{eq:window1}) the fluxes are chosen. With these results and the change of variables (\ref{eq:change-var}), we obtain the time evolution of the original fields $g_\text{s}$, $L_1$ and $L_2$,
\be\label{eq:solmodel2}
g_\text{s} \sim e^{16k \sigma_1} \left(\frac{K^2}{V_* t^2}\right)^{32k^2},\;\;L_1 \approx L_2 \sim e^{\frac{\sigma_1}{16k^2}} \left(\frac{K^2}{V_* t^2}\right)^{\frac{1}{32k}},\;\frac{g_\text{s}^2}{L^n} \sim e^{-2 \sigma_1}\,\frac{K^2}{V_* t^2}\,.
\ee
As before,
the time-dependence of the effective coupling follows from (\ref{eq:scaling-geff}), and obtaining it directly from $g_s$ and $L_i$ requires keeping the exact coefficients presented above in (\ref{eq:change-var}).
We expect similar sequences of models on other Ricci-flat spaces, although it would be more difficult to analyze them as explicitly.

We will analyze the behavior of the string coupling and radii in this model below in \S\ref{subsec:control}, finding that we need to lift a significant fraction of the $2^D$ Ramond-Ramond axion fields in order to obtain a controlled perturbative expansion.  In order to achieve this, we will introduce two additional ingredients:  Neveu-Schwarz flux and topology for it to thread.  These elements will complete the model in a way that leaves the potential (\ref{eq:Veff2}) and the resulting solution with acceleration (\ref{Ktoroidal}) as a good approximation, but with the additional internal topology the model will no longer be a toroidal orientifold.    

\subsection{Structure and control of the solutions}\label{subsec:control}

In this subsection we will analyze the parameters in our solution and their perturbative control. We will focus on the limit of large $k$, for which $K \gg 1$; de Sitter is obtained to very good approximation for $k \to \infty$ as the equation of state approaches $w=-1$.  This regime is at large total dimension $D$, where various interesting effects and simplifications arise.  

\subsubsection{General requirements}

We must arrange the different contributions to the potential so that the radii are large in string (and Planck) units, the coupling is weak, and the time evolution of the scalars and metric is controllably small relative to the string scale.  

There are various important effects that arise in the regime of large $D$.  First, there is a large number of RR fields:  of order $2^D\sim 2^n$ \cite{SCdS}.  On a compactification down to $d=4$ these include of order $2^n$ axions, along with higher harmonics on the internal space.  For our analysis of control it is important to consider two regimes of energy scales:  those below the compactification scale $1/L$  and those above $1/L$.  

At scales below $1/L$, the interactions are controlled by the effective coupling
\beq\label{geff}
g_\text{eff} = \frac{g_\text{s}}{L^{n/2}}
\eeq
with an enhancement from the number $N_\text{light}$ of light species. 
If this were a large effect, it would make an interesting moduli-dependent modification of the potential.  In particular, it would renormalize the Planck mass, and change the form of the factor one obtains in converting to Einstein frame in $d$ dimensions as a function of $L$, $g_\text{s}$, and other moduli -- this may in particular modify the classical runaway behavior near large volume.  It will be interesting to explore the implications of this new structure for moduli stabilization and dynamics, since this large number of axion fields is a striking feature of the string spectrum. 

However, in the present work we will require for simplicity that the tree level model satisfies
\beq\label{geffcond}
N_\text{light}\,g_\text{eff}^2  \ll 1
\eeq  
for our solutions to ensure that the dynamics developed above is a good approximation.  In our sequence of models discussed in \S\ref{subsec:tori}\ which approach $w\to -1$, we will need to introduce additional ingredients, subleading in the classical dynamics,  to lift the $\sim 2^D$ axion fields.  In our sequence of models with $K\gtrsim 1$, we will find (\ref{geffcond}) to be satisfied with $N_\text{light}\sim 2^D$, so in that case we will not require any further ingredients.   

At scales above $1/L$, interactions are controlled by $g_\text{s}$ directly, with a species enhancement factor.  However, in this regime we also have important $D$-dependent suppression factors from the angular part of loop momentum integrals \cite{StromingerD}, which contributes
\beq\label{Omegaloop}
C\equiv \int \frac{\d\Omega_{D}}{(2\pi)^D} = \frac{2^{-D+1}}{(2\pi)^{D/2}\Gamma(\frac{D}{2})},
\eeq      
with the radial part of the momentum integrals effectively cut off at the string scale.  
This factor $C$ generalizes the loop suppression factor of $16\pi^2$ in familiar four-dimensional perturbation theory.  
Since powers of $g_\text{s}^2$ count loops, with of order $2^D$ species running in the loops, we impose that
\beq\label{gcondD}
2^D g_\text{s}^2 \ll \frac{1}{C}.
\eeq
A similar decoupling happens in classical large-$D$ general relativity at a fixed value of the gravitational coupling $\kappa$:  the Newtonian potential between sources is negligible outside a very small radius \cite{EmparanD}.  In the work \cite{StromingerD}\ on large-$D$ perturbative quantum general relativity, $\kappa$ was rescaled by the inverse of the phase space factor  $C$ (\ref{Omegaloop}) in order to obtain surviving loop corrections in the large-D limit.  Here, we work at large but finite $D$ and will show that our solutions for $g_\text{s}$ satisfy (\ref{gcondD}). In fact, the model of \S \ref{subsec:simpletori} will be seen to satisfy the stronger inequality $g_s \ll 1$.

Having analyzed the couplings and radii, let us next consider the scale of the curvature and scalar time dependence in our solutions.  First, recall from \cite{BP} that at least if we start from a sufficiently general distribution of initial radii, we can tune the minimum of the effective potential for $\sigma$; i.e. we can tune $V_*$ in (\ref{Vstar}).  
To see the utility of this feature in our explicit models,
let us analyze the level time-dependence compared to the Planck and string scales. 
Since $a(t)=(t/t_0)^K$, we have that the $d$-dimensional Hubble parameter (in Einstein frame) is given by 
\beq\label{HEinstein}
H_{\text{Einstein}}=\frac{\dot a}{a}=\frac{K}{t}
\eeq
This dies to zero at late times, and hence is much smaller than $ M_d$ at sufficiently late times.    
But we would like to also determine the level of $\alpha'$ corrections generated by the curvature and scalar time derivatives in our solution.  For this, it is useful to return
to the string-frame metric  (\ref{metrics}), which we can write as
\bea\label{metricsagain}
ds^2_{D,\text{str}}&=&g_{\text{eff},0}^{\frac{4}{d-2}}\left(\frac{t_0}{t}\right)^2\left(-dt^2+\left(\frac{t}{t_0}\right)^{2K}d\vec x^2\right) + L^2\gamma_{ij} dy^i dy^j \\ \nonumber
& &\\ \nonumber
&=&-d\tau^2+a(\tau)^2 d\vec x^2 + L^2\gamma_{ij} dy^i dy^j\nonumber
\eea
with
\be\label{astring}
 a(\tau) = g_{\text{eff},0}^{\frac{2}{d-2}}\exp\left((K-1)\frac{\tau-\tau_0}{t_0g_{\text{eff},0}^{\frac{2}{d-2}}}\right)
\ee
where $g_{\text{eff},0}$ is the effective low energy $d$-dimensional string coupling $g_\text{s}/L^{n/2}$ evaluated at time $t=t_0$.  Here we used the time evolution of the effective coupling given in (\ref{eq:scaling-geff}).
This reduces to the original supercritical linear dilaton solution for $K=1$, for which the string-frame metric is Minkowski spacetime.
In our more general solutions with $K>1$, the string-frame metric undergoes accelerated expansion (as does the $d$-dimensional Einstein frame metric).  

The curvature and scalar time derivatives in string units are given by
\beq\label{Hstrings}
\frac{da/d\tau}{a}=\frac{K-1}{t_0g_{\text{eff},0}^{\frac{2}{d-2}}}\;, ~~~\frac{dL/d\tau}{L}=\frac{t}{t_0g_{\text{eff},0}^{\frac{2}{d-2}}}\frac{\dot L}{L}\;, ~~~~\frac{dg_\text{s}/d\tau}{g_\text{s}}=\frac{t}{t_0g_{\text{eff},0}^{\frac{2}{d-2}}}\frac{\dot g_\text{s}}{g_\text{s}}\,.
\eeq
In order to make these controllably small, we can tune $V_*$ to be small, hence increasing $t_0$ while maintaining the solution (\ref{Vstar}) of the rolling scalar equation of motion. This limit is simpler in some ways than the supercritical linear dilaton solutions, in that the curvature and scalar field time derivatives can be below the string mass scale (whereas in the linear dilaton solution the time dependence is of order the string scale, and one controls $\alpha'$ corrections by using the exact worldsheet solution).

Finally, we need to check that corrections from localized sources to our effective theory are negligible. In our case, these arise predominantly from the orientifold planes which, unlike the supercritical potential or the RR fluxes, are localized in the internal directions. The effective potential that we have used so far includes the average of the orientifold tension over the internal space, and corrections from the localization of such sources appear in the form of gradients $\int (\nabla A)^2$ of the warp factor $A(y)$ \cite{Douglaswarped}. As shown in \cite{micromanaging}, these effects can be neglected if $A \ll 1$ away from the cores of the O-planes.

Next we will implement these conditions for control in the two sequences of models developed above in \S\ref{subsec:simpletori}-\ref{subsec:tori}.  

\subsubsection{The models of \S\ref{subsec:simpletori}}

We now analyze the conditions under which the models of \S\ref{subsec:simpletori} are under perturbative control. From the time dependence (\ref{eq:solmodel1}), we see that $g_\text{s}$ and $L$ are increasing with time -- although very slowly at large $k$.  As a result, we will find only a finite window of times during which the solution is under control.  To check that there is a large window of times for which the solution applies, we need to assess the values of the coupling and radii at some initial time $t_0$, and establish that the conditions for a controlled expansion are satisfied parametrically.    
The fact that the window of times is finite is reminiscent of the fact that de Sitter solutions in string theory decay.  It is somewhat intriguing in that our models provide another, more perturbative, context where accelerated expansion occurs, but not indefinitely.  

We now check that the control conditions
\begin{equation}
L_1 = L_2^2 \gg 1 \,,\qquad
2^n \frac{g_s^2}{L^n} \ll 1 \,,\qquad
g_s \ll 1 \,,
\end{equation}
are satisfied. (Note that we will satisfy the stronger $g_s \ll 1$, instead of  \eqref{gcondD}). For this, it is useful to eliminate $\Phi$ in terms of $L_2$.
Since $g_s/L_2^{9k} \sim 2^{-\frac n4}$ and $n=6k^2$, we find that the control window is
\begin{equation}
2^{\frac{k}{3(k-2)}} \ll L_2 \ll 2^{\frac k6} \,.
\end{equation}
This can be translated into a finite window of times by use of (\ref{eq:relation1}) and $e^\Phi=K(3K-1)/V_*t^2$.

We also need to check the smallness of time derivatives (\ref{Hstrings}) in string units. The gradients for the scale factor and $L$ are proportional to $V_*^{1/2}/k^2$, while the time derivative of $\log\,g_\text{s}$ is proportional to $V_*^{1/2}/k$. These gradients are time-independent, and can be made small by taking $k$ large and/or tuning $\epsilon \ll 1$ in (\ref{eq:sigma1Vstar}). 

Lastly, let us check that the warp factor caused by the localized O-planes is small. There are a total of $n_1+n_2$ directions transverse to each O-plane.  Of these there are $n_1$ directions that have size $L_1$, and the remaining $n_2$ directions have size $L_2$.  Let us denote these directions respectively by $\vec y_1$ and $\vec y_2$.  Schematically we have
\begin{equation}
\nabla^2 A(\vec y_1, \vec y_2) \sim g_s^2 \frac{1}{g_s} \sum_{\vec m_1, \vec m_2} \delta^{(n_1)}(\vec y_1 -\vec m_1 L_1) \delta^{(n_2)}(\vec y_2 -\vec m_2 L_2) +\text{other sources} \,,
\end{equation}
where the first factor $g_s^2$ arises from Newton's constant, the second factor $1/g_s$ is the O-plane tension, and we sum over a periodic array of images of the O-plane which we have conveniently put at $\vec y_1=\vec y_2=0$.

The warp factor $A$ halfway in the middle of the O-plane and its images is
\begin{multline}\label{eq:warp1}
A\(\frac{L_1}{2} \vec1, \frac{L_2}{2} \vec1\) \sim g_s \sum_{\vec m_1, \vec m_2} \frac{1}{\[L_1^2\(\vec m_1-\frac12\vec 1\)^2 +L_2^2\(\vec m_2-\frac12\vec 1\)^2\]^{(n_1+n_2-2)/2}} \\
+\text{contributions from other sources} \,,
\end{multline}
where $\vec 1$ denotes the vector $(1,1,\cdots,1)$.  As usual, the contributions from other sources are such that $A$ is the the difference between the sum and integral over $\vec m_1$, $\vec m_2$.  Since our solution gives $L_1=L_2^2$, a conservative estimate for $A$ that gives an upper bound is
\begin{equation}
A\(\frac{L_1}{2} \vec1, \frac{L_2}{2} \vec1\) < \frac{g_s}{L_2^{n_1+n_2-2}} \sim 2^{-\frac n4} L_2^{3k+2} = 2^{-\frac{3k^2}{2}} L_2^{3k+2} \,.
\end{equation}
This is small during our entire window of control.  Note that it is not important to include factors such as $\Gamma\(\frac{n_1+n_2}{2}\) \sim k^{3k}$ which may arise from the sum or integral over the $n_1+n_2$ transverse coordinates $\vec y_1$, $\vec y_2$.

In summary, for $k \gg 1$ this sequence of models is perturbative for a parametrically large window of times.

\subsubsection{The models of \S\ref{subsec:tori}}

Our second sequence of models will be more subtle and require some modifications to treat the effects of the many axions, while retaining the original solution for the radii and couplings as a good approximation.

In general there are about $2^n$ RR axions which may pose a light species problem; indeed one finds as we will see that in contrast to the previous sequence of models, here we cannot satisfy \eqref{geffcond}\ with $2^n$ light axions contributing to $N_\text{light}$.  Because of that, we will introduce $H_3$ fluxes to lift (most of) the RR axions, using the terms in the Lagrangian of the form \cite{joebook}
\beq\label{tildeF}
|\tilde F_{r+3}|^2 =|F_{r+3}+C_r\wedge H_3|^2 .
\eeq
There are two types of such terms, (i) ones for which we have prescribed a background flux, i.e. $r+3=m(p_{11}+p_{12})$ or $r+3=m(p_{21}+p_{22})$, and (ii) other values of $r$.  In case (ii), the term \eqref{tildeF}\ gives a mass for the axions.   

These fluxes also introduce further D-brane charge tadpoles, generalizing those in \cite{GKP}\cite{KKLT}.  D-branes can cancel these charges without introducing a leading contribution to the potential energy.  

In case (i) the axion $C_r$ is up on a potential hill; we will find a consistent regime of axion monodromy inflation \cite{monodromy}\ in those directions.  The potential for the canonically normalized axion field $\phi_\text{c}$ is of the form
\beq\label{largephi}
m^2(\phi_\text{c}-\phi_{\text{c}0})^2
\eeq
where $m$ and $\phi_0$ depend on the moduli.  Below we will impose the condition for slow roll in the $\phi_\text{c}$ direction, ensuring that it is slower than the rolling field $\Phi$ in our original solution.  As standard in large-field inflation, this occurs for sufficiently large values of $(\phi_\text{c}-\phi_{\text{c}0})/M_P$.
This mechanism for inflation at large field values \cite{monodromy}\cite{inflationreviews}\ was realized and developed after the first round of moduli stabilization efforts \cite{landscapereviews}.  In the original examples realizing the mechanism, this was set up within those earlier moduli stabilization scenarios, with the inflationary mechanism built in as a module.  Here we will see that it can participate in new stabilization and acceleration mechanisms in a less modular, more economical way as we will see.   

\begin{figure}[h!]
\begin{center}
\includegraphics[width=0.4\textwidth]{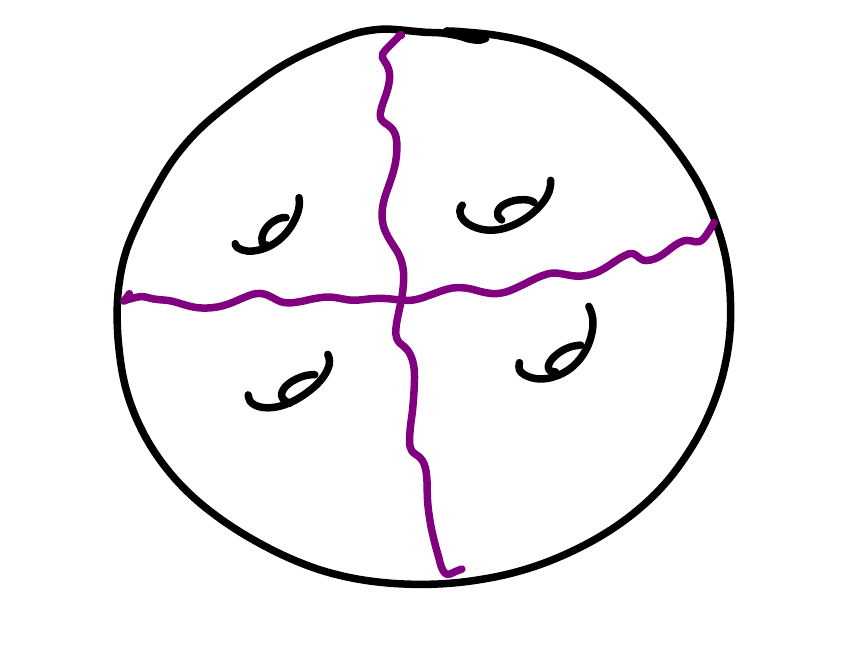}
\end{center}
\caption{In order to preserve sufficient fluxes to lift our axions, we consider
a geometry which generalizes the toroidal model discussed earlier. Here, the orientifold planes (purple)
act freely on the homology cycles, leaving invariant fluxes on odd or even combinations of them.}\label{fig:Ofigure}
\end{figure}

In our setup of \S\ref{subsec:tori}, many of the $H_3$ flux choices on the $n$-torus are projected out by the orientifolds.  Because of that, we will consider a more general geometry than a torus, for example a product of Riemann surfaces, and place the orientifolds so that they do not fix the nontrivial 1-cycles of the Riemann surfaces (see Figure \ref{fig:Ofigure}).  As mentioned above, this means that flux on these cycles is projected in by the orientifold, in either an even or odd combination.  The contribution to the $4d$ potential energy from the internal curvature is subdominant to our leading terms in the potential, including the classical supercritical potential, since the latter goes like $D-D_\text{crit}$ where as the former goes like the inverse curvature radius squared, $\sim 1/L^2\ll 1$. 

Next, we will analyze the contribution of $H_3$ flux to the axion masses (to make sure they are large enough) and to the potential (insisting that this be subdominant to the original terms used in our scaling solution).  These conditions, along with the condition \eqref{gcondD}\ will be the leading constraints on our parameters, as we will see after assessing all of the requirements.  

Let us first consider the simple case where $H_3$ is completely symmetrized across all internal dimensions.  There are of order $n^3$ three-cycles in the internal manifold, and let us denote the $H_3$ flux number on each particular three-cycle as $N_3$.  First, we would like to keep the total potential energy from $H_3$ parametrically smaller than other potential terms that we have considered.  Comparing it to the supercritical term, we get
\be\label{h3cond}
n^3 \frac{N_3^2}{L^6} \ll (D-D_\text{crit}) \sim n \quad
\Rightarrow \quad L \gg (N_3 n)^{1/3} \,.
\ee
Note that this is strictly stronger than the condition $L\gg1$.

Next, we analyze how $H_3$ affects the masses of RR axions.  These axions come from the zero mode of $C_r$, and gain masses from the coupling $|C_r \wedge H_3|^2$.
So each time this occurs, the mass squared of a particular $C_r$ gains $N_3^2/L^6$, which can seen by comparing the quadratic potential $|C_r \wedge H_3|^2$ with the kinetic term $|dC_r|^2$.
To lift an axion we need to increase its mass squared to at least $1/L^2$.  For any $C_r$ with $r\le n-3$ there are of order $n-r \choose 3$ $H_3$ fluxes with nonzero wedge product $C_r \wedge H_3$.  Let us choose a threshold $r_0=n-n^{2/3}$ and lift all $C_r$ with $r\le r_0$.  This means
\be\label{m2cond}
{n-r_0 \choose 3}\frac{N_3^2}{L^6} \gg \frac{1}{L^2} \quad
\Rightarrow \quad L \ll (N_3n)^{1/2} \,.
\ee

There are at most of order $(n-r_0){n \choose r_0} \sim n^{n^{2/3}}$ axions that do not get lifted this way -- these are $C_r$ with $r>r_0$.  Therefore we have reduced the number of light species from $2^n$ to at most $n^{n^{2/3}}$.  We would like to make sure that the effective coupling (after enhanced by the species) is small:
\be\label{newsp}
n^{n^{2/3}} \frac{g_\text{s}^2}{L^n} \ll 1 \,.
\ee
We also must satisfy the weak coupling condition \eqref{gcondD}\ arising from scales above $1/L$, which requires
\be\label{eq:gsconstr}
L \ll n^{1/2}\,.
\ee

Next let us analyze the axions which are away from their minimum in our solution,
those which have a tadpole via the coupling $|F_p + C_{p-3} \wedge H_3|^2$ where $F_p$ is the background RR flux in our model.  
We can satisfy the slow roll condition by requiring a large canonical axion field $\phi_\text{c}\gg M_P$ \eqref{largephi}.  For concreteness let us first consider the background flux wrapping $L_1$ with $p_1=m(p_{11}+p_{12})$.  In terms of the canonical axion field
\be
\phi_{c_1} \sim L^{n/2}\frac{c_1}{L_1^{mp_{11}-3}L_2^{mp_{12}}}
\ee
we can write the effective potential from $|F_{p-1} + C_{p_1-3} \wedge H_3|^2$ as
\be
V_\eff \sim L^n \frac{(Q_1+c_1 N_3)^2}{L_1^{2mp_{11}}L_2^{2mp_{12}}} \sim \frac{L^n}{L_1^{2mp_{11}}L_2^{2mp_{12}}} \(Q_1 + \frac{L_1^{mp_{11}-3}L_2^{mp_{12}}}{L^{n/2}} \phi_{c_1} N_3 \)^2 \,.
\ee
Here for simplicity we have chosen $H_3$ to be along the $L_1$ directions, but this does not affect the final result at leading order.  We have a large number of axions from the RR potential fields $C_{p-3}$, the number being of order $n^3$, so the situation is similar to N-flation \cite{Nflation}.  
In particular, the effective slow roll parameter in the collective direction is suppressed by $1/\sqrt{N_{\rm inflaton}}\sim n^{-3/2}$:
\be
\varepsilon = n^{-3/2} M_4\frac{|\partial_{\phi_{c_1}} V_\eff|}{V_\eff} \sim n^{-3/2} \frac{L^{n/2}}{g_\text{s}} \frac{N_3 L_1^{mp_{11}-3}L_2^{mp_{12}}}{Q_1 L^{n/2}} = \frac{N_3 L_1^{mp_{11}-3}L_2^{mp_{12}}}{n^{3/2} g_\text{s} Q_1} \,.
\ee
Let us require this slow roll parameter to be much smaller than $\beta_\text{c}\sim \sqrt{1/K}\sim k^{-1/2}$, so that corrections to our previous solution can be neglected.  This means
\be\label{q1cond}
\frac{g_\text{s} Q_1}{L_1^{mp_{11}-3}L_2^{mp_{12}}} \gg \frac{N_3 k^{1/2}}{n^{3/2}} \sim \frac{N_3}{n^{4/3}} \,,
\ee
and similarly for the background flux wrapping $L_2$ we have
\be\label{q2cond}
\frac{g_\text{s} Q_2}{L_1^{mp_{21}}L_2^{mp_{22}-3}} \gg \frac{N_3}{n^{4/3}} \,,
\ee

We are now ready to solve these conditions by plugging in our solution. The calculation of the window of control is a bit more subtle than in the previous model, due to large cancellations when $k \gg1$. It turns out to be convenient to first derive a parametric window in terms of $L$, and then translate this into a window of times using (\ref{eq:solmodel2}). Eliminating $\Phi$ in favor of $L$ from $L \sim e^{\frac{1}{32k} \Phi + \frac{1}{64k^2} \sigma_1}$, obtains\footnote{Some of these results appear at subleading order in an expansion in $1/k$, so it is best to calculate the various combinations exactly and only take the large-$k$ limit at the end. Also, recall that $\sigma_2 \approx 0$ at the minimum.}
\ba
g_\text{s} &\sim L^{1024k^3} e^{-\sigma_1}\;,\;\;
\frac{g_\text{s}}{L^{n/2}}  \sim L^{16k-3} e^{-\sigma_1} \\
\frac{g_\text{s}}{L_1^{mp_{11}-3}L_2^{mp_{12}}} &\sim \frac{g_\text{s}}{L_1^{mp_{21}}L_2^{mp_{22}-3}}\sim  L^3 e^{-\sigma_1}\,. \nonumber
\ea
We also recall that in this class of models $n\approx2048 k^3$.

Using the scalings at the minimum $e^{\sigma_1}\sim Q_1 \sim Q_2 \sim |V_\text{O}| \sim 2^{n/4}$, we find that the weak effective coupling condition \eqref{newsp} reduces to
\be
n^{n^{2/3}} L^{32k} \ll 2^{n/2} \quad
\Rightarrow \quad L \ll 2^{32k^2} \,,
\ee
whereas the slow roll conditions \eqref{q1cond} and \eqref{q2cond} become
\be
L^3 \gg \frac{N_3}{n^{4/3}} \quad
\Rightarrow \quad L \gg N_3^{1/3} n^{-4/9} \,.
\ee
Even without the N-flation effect we get $L\gg N_3^{1/3} n^{1/18}$.  Combining these inequalities with \eqref{h3cond} and (\ref{eq:gsconstr}) (\eqref{m2cond} gives a weaker upper bound), we arrive at
\be\label{eq:windowtemp}
(N_3 n)^{1/3} \ll L \ll n^{1/2} \,,
\ee
which can be satisfied within a parametrically large window as long as $N_3 \ll n^{1/2}$. 

On the other hand, the time derivatives (\ref{Hstrings}) are proportional to $(2^{n/2} V_*)^{1/2}$, so by tuning $V_* \ll 2^{-n/2}$ the corrections from gradients are negligible. From (\ref{eq:sigmamodel2}), $V_*$ is already suppressed by $|V_\text{O}|^2 \sim 2^{-n/2}$, so the required tuning for $\epsilon$ in (\ref{eq:sigmamodel2}) is power-law in $n$. Finally, following steps similar to those around (\ref{eq:warp1}), one may verify that the warp factor $A \ll 1$, so that corrections from the localization of O-planes can be self-consistently neglected.
In summary, we find that the sequence of models approaching $w \to -1$ is under perturbative control for a parametrically large window of times
\be
\frac{2^{1024 k^2}}{n^{16k}} \ll V_* t^2 \ll \frac{2^{1024 k^2}}{(N_3n)^{32k/3}}\,,
\ee
with $N_3 \ll n^{1/2}$ and $n=2048 k^3$.

In summary, in this section we have shown that an elaborated version of the models of \S\ref{subsec:tori}\ satisfies the various consistency requirements for control.  The leading ones turned out to be the condition that most of the axions are lifted above the scale $1/L$, that the fluxes introduced to accomplish this do not contribute a leading effect to the moduli potential, and that the microscopic string coupling remain sufficiently small.  

\section{Accelerated expansion at finite density}\label{sec:density}

There are other broad classes of mechanisms for generating accelerated expansion in which densities of particles, strings, or higher dimensional defects make leading contributions to the stress-energy \cite{thermal,trapped}.  For example, from the bottom up domain wall networks constitute a fluid with equation of state parameter $w=-2/3$, equivalently $K=2$.      
In string theory, it is not as simple as that -- the energy density carried by such sources depends also on moduli fields, which can evolve in time.  

In this section we present a class of examples producing accelerated expansion taking this into account, with the examples we will derive below producing specific values of $K$ in the range
\beq\label{Krho}
1 < K < 2\,.
\eeq
In these models, the expansion is sourced entirely by microscopically consistent domain walls.  We will work here for simplicity in $D=10$ on a torus and obtain a finite list of models, but we expect that such effects may assist inflation much more generally.  

This method provides a new way to simplify the construction of accelerating cosmologies, limiting number of ingredients required.  One application is to holography.  After explaining the models, we will comment on holographic examples and their interpretation below.   

At a technical level, the distinction is the following.  The models that we presented in the previous sections are constructed out of objects that wrap the entire non-compact spacetime. Of course, this is not the only possibility: we may choose to wrap any number of the non-compact dimensions, leading to a density of extended or point-like objects in spacetime. In order to preserve isotropy in both the internal space and the $d$-dimensional spacetime, we will symmetrize the objects over all of the directions. The effective picture in the FRW spacetime is then an isotropic network of extended objects (or cosmic strings in $d=3$). For a specific class of models, this turns out to lead to accelerating solutions without tachyonic moduli.
\subsection{Single-term potentials}\label{subsec:singleterm}
\indent In string theory, we have two classes of extended objects to choose from. The first consists of D-branes, which have tension $1/g_{\text{s}}$. There also exist heavier objects, such as NS5-branes and $(p,q)$ 7-branes, whose tensions go as $1/g_{\text{s}}^2$. Fixing the critical dimension for simplicity, the spacetime effective potentials\footnote{By which we mean the energy density.} for these two types of objects are
\bea\label{denseV}
V_{\text{D}-\text{brane}}&=&\frac{\rho}{a(t)^{d_\perp}}e^{\left(\frac{d+2}{2}-d_\perp\right)\tilde{D}+\left(p+d_\perp-\frac{d}{2}-4\right)\phi}\\
V_{\text{heavy}}&=&\frac{\rho}{a(t)^{d_\perp}}e^{(2-d_\perp)\tilde{D}+(p+d_\perp-9)\phi} \nonumber\,.
\eea
Here $d_\perp$ is the codimension of the brane in spacetime, $\rho$ is the density of objects, and $p$ is the total number of spatial dimensions on the brane. The sources are smeared over all spatial dimensions, so that the solution is isotropic.

In order to find a scaling solution, it is convenient to consider a slightly more general potential,
\be
V=\frac{\rho}{a(t)^{d_\perp}}e^{\alpha\phi_1+\beta \phi_2}\,,
\ee
where $\phi_1$ and $\phi_2$ are canonically normalized. 
Since we are dealing with single-term potentials, there is only one linear combination of fields $\Phi$ that is sourced by the density. The combination $\sigma$ orthogonal to this is a flat direction in the classical supergravity approximation, which is presumably lifted at higher order in the string coupling. More specifically, the two canonically normalized scalars are 
\be\label{eq:newchange}
\Phi=\frac{\alpha \phi_1+\beta \phi_2}{\sqrt{\alpha^2+\beta^2}}\;,\;\;\sigma=\frac{\beta \phi_1-\alpha \phi_2}{\sqrt{\alpha^2+\beta^2}}\,.
\ee
The potential is now simply
\be
V=\frac{\rho}{a(t)^{d_\perp}}e^{\sqrt{\alpha^2+\beta^2}\Phi}\,.
\ee

Next we plug this into the equations of motion (\ref{eom}), along with a FRW ansatz with scale factor $a=(t/t_0)^K$. One finds a scaling solution, with
\bea\label{eq:densitysol}
\Phi&=&\Phi_0-\frac{2-d_\perp K}{\sqrt{\alpha^2+\beta^2}}\log \left(\frac{t}{t_0}\right)\\
K&=&\frac{2(2d-d_\perp-2)}{(d-1)(d-2)(\alpha^2+\beta^2)+2(d-1)d_\perp-d_\perp^2}\nonumber\,,
\eea
and $\Phi_0$ is determined in terms of the density $\rho$ by
\be\label{eq:rho}
e^{\sqrt{\alpha^2+\beta^2} \Phi_0} t_0^{2-d_\perp K}= \frac{\rho^{-1}}{\alpha^2+\beta^2}(2-d_\perp K)\left((d-1)K-1\right)\,.
\ee
For $d \ge 2$, these results give an upper bound on $K$,
\be
K \le \frac{2}{d_\perp}\,,
\ee
which is saturated by a perfect fluid. In order to find accelerated expansion, we will then focus on domain walls, $d_\perp=1$.

We now apply this macroscopic analysis to the case of our microscopic potentials from string theory. The solution is
\bea\label{Kheavy}
K_{\text{D}-\text{brane}}&=&\frac{2(2d-d_\perp-2)}{2(d-1)d_\perp-d_\perp^2+(d-1)(d-2)\left(\frac{(d+2-2d_\perp)^2}{4(d-2)}+\frac{\left(p+d_\perp-\frac{d}{2}-4\right)^2}{10-d}\right)}\\
K_{\text{heavy}}&=&\frac{2(2d-d_\perp-2)}{2d_\perp(d-1)-d_\perp^2+(d-1)(d-2)\left(\frac{(d_\perp-2)^2}{d-2}+\frac{(d_\perp+p-9)^2}{10-d}\right)}\nonumber\,.
\eea
Scanning through the values of $d_\perp$, $d$, and $p$, one finds that D-branes and NS5 branes do not lead to accelerated expansion. On the other hand, for the case of $(p,q)$ 7-branes with codimension one, we find
\be
K=2-\frac{16 (d-1)}{d (31-2d)-38}\,,
\ee
which gives three accelerating models, $(d, K)= (3,\frac{42}{37}),\,(4,\frac{10}{9}),\,(5,\frac{70}{67})$.
In all three cases, $\rho$ is positive and is related to $\Phi_0$ and $t_0$ by (\ref{eq:rho}). It is perhaps intuitive that $(p,q)$ 7-branes are the most likely to give acceleration, since they are the heaviest and biggest objects in string theory.

As in the supercritical models presented above, it is necessary to check that there is a window of time where our solutions are under perturbative control. Inverting (\ref{eq:newchange}) and using (\ref{eq:densitysol}), obtains
\be\label{timedep}
g_{\text{s}}=g_{\text{s}0}\,,\;\;L=L_0\left(\frac{t}{t_0}\right)^{\frac{1}{8}(d-2)(2- d_\perp K)}\,,
\ee
and the effective coupling,
\be
g_\text{eff}^2=\frac{g_\text{s}^2}{L^n}=\frac{g_{\text{s}0}^2}{L_0^n}\left(\frac{t}{t_0}\right)^{-\frac{1}{8}n(d-2)(2- d_\perp K)}\,.
\ee
Here, the number of internal dimensions is $n=10-d$. Note also that $g_\text{s}$ refers to the average string coupling, since the axio-dilaton is not spatially constant, and undergoes a monodromy around the $(p,q)$-7 branes. In these accelerating solutions, the string dilaton is a flat direction and, recalling that $K< 2/d_\perp$, $L$ grows and $g_\text{eff}^2$ decreases with time. It follows that the solutions are under control at sufficiently late times.

Finally, we check that there are no large gradients. In string frame, $d \tau = g_\text{eff}^{2/(d-2)} dt$, the gradients $ d \log a/ d\tau$ and $d \log L / d\tau$ are proportional to $g_\text{eff}^{-2/(d-2)}/t$, which decreases with time.
Therefore, $\alpha'$ corrections become unimportant at sufficiently late times.

\subsection{Constructing a stable domain wall network}

There is an implicit assumption in the above analysis that is crucial to its consistency: we have taken  for granted that there are no perturbative instabilities in the brane network. Our solutions would not be valid if there were a tachyon between two intersecting branes, since then the network would evolve in a more complicated way, and the energy density would no longer take the simple form (\ref{denseV}). Another source of potential instabilities is motion collective coordinates of the branes.  In this section, we will give an argument for the existence of perturbatively stable networks of domain walls in the case $d=4$.

The general strategy is to isotropize the branes in the internal and noncompact spaces, while making sure that intersecting branes have no tachyon. Although the full FRW solution is not supersymmetric, we may eliminate tachyons by requiring that branes that intersect be mutually supersymmetric, since this condition implies that the force between the branes vanishes. For two orthogonally intersecting branes, one quarter of the supersymmetries are preserved if
\begin{align}
\#_{\text{ND}}=4,
\end{align}
where $\#_{\text{ND}}$ is the number of directions that is orthogonal to one of the branes and parallel to the other. This constraint on the network becomes increasingly stringent as the codimension of the branes decreases, since they become more likely to intersect. In particular, domain walls must intersect unless they are parallel.

In the case $d=4$, let us consider two classes of three $(p,q)$-7 branes. The first class is oriented as 
 \begin{align}
 \begin{tabular}{c|cccc|cccccc}
 & 0 & 1 & 2 & 3 & 4  & 5 & 6 & 7&8&9\\
\hline
I &X &  &  X& X& &X &X &X &X  &X \\
II&X & X& &X & X& & X& X& X &X\\
III &X & X&X & & X& X& &X & X &X\\
\end{tabular}
 \end{align}
where $0,1,2,3$ are coordinates on the FRW spacetime. Similarly, the branes in the second class have orientations
\begin{align}
\begin{tabular}{c|cccc|cccccc}
 &0 & 1 & 2 & 3 & 4  & 5 & 6 & 7&8&9\\
\hline
I'& X &  &  X& X& X&X &X &&X  &X \\
II'&X & X& &X & X& X& X& X&  &X\\
III'&X & X&X & & X& X& X&X & X \\
\end{tabular}
 \end{align}
 It is clear that the branes in each class are mutually supersymmetric among themselves. However, the branes in the first class are not mutually supersymmetric with their primed partners in the second class, so we must keep them at finite separation in the FRW space. In particular, the density $\rho$ cannot be too large in order to lift the tachyon between the branes. Alternating between the two classes in each noncompact dimension then leads to a network free of tachyons from $(p,q)$ strings between the branes.  Moreover, because the attractive potential between the separated branes is linear, there is no tachyon from the motion collective coordinates.
 
In $d=3$ and $d=5$ it seems more difficult to construct such a network, due to the odd number of internal dimensions. We suspect that this may be possible using branes oriented at angles and/or other internal geometries, but we will not attempt to do so here. Another possibility is to relax the assumption of isotropy of the internal space; it is not hard to check that doing so leads to an accelerating and stable network in $d=3$.

 \subsection{Two-term potentials}
 
 Let us now explore a more general case where the potential is the sum of two positive terms,
 \begin{align}\label{twoterm}
 V=\frac{\rho_1}{a(t)^{d_{\perp1}}}e^{\alpha_1\phi_1+\beta_1\phi_2}+\frac{\rho_2}{a(t)^{d_{\perp2}}}e^{\alpha_2\phi_1+\beta_2\phi_2}.
 \end{align}
As in the single-term case, there is one linear combination $\Phi$ of fields that rolls with time, and another linear combination $\sigma$ that remains constant. There are a plethora of accelerating solutions of this form in string theory, and we will now consider some interesting examples.  We will not analyze them in detail here; in particular the question of perturbative stability of the scalar fields and the network is more subtle in this case.

First, suppose that we try to add a density of NS5-branes to the solutions with $(p,q)$-7s that we found above. The resulting potential is 
\begin{align}
V=\frac{\rho_{7}}{a(t)}e^{\tilde{D}-\phi}+\frac{\rho_5}{a(t)^{d_{\perp}}}e^{(2-d_{\perp})\tilde{D}+(d_{\perp}-4)\phi}.
\end{align}
Plugging this into the equations of motion gives a solution in $d=3$ with
\begin{align}
K=1+\frac{2-d_{\perp}}{17+4d_{\perp}(2d_{\perp}-5)}.
\end{align}
It follows that the model accelerates for $d_{\perp}=0$, which is an NS5 brane wrapped on the entire noncompact space. In fact, this is the unique accelerating model that we have found with both NS5 branes and $(p,q)$-7s.\\
\indent We can also try to add lighter objects to our original solutions. For example, let us consider codimension-zero D$p$-branes in Type IIB. The potential becomes
\begin{align}
V=\frac{\rho_{7}}{a(t)}e^{\tilde{D}-\phi}+V_{\text{D}p}e^{\frac{d+2}{2}\tilde{D}+\left(p-\frac{d}{2}-4\right)\phi}.
\end{align}
Again there is a scaling solution, whose exponent in $d=4$ is 
\begin{align}
K=\frac{174+p(5p-49)}{126+4p(p-9)}.
\end{align}
This yields acceleration for the case of D7-branes (for all the others, the solution requires a physically unacceptable negative density $\rho<0$). The resulting $K$ is suppressed compared to the original solution with no D-branes.

Although we checked above that NS5 branes do not lead to acceleration by themselves, one might wonder whether they could yield acceleration when combined with other ingredients. This indeed turns out to be the case. For instance, let us consider a model with a density of NS5 branes and Ramond-Ramond flux $F_p$,
\begin{align}
V=\frac{\rho_5}{a(t)^{d_{\perp}}}e^{(2-d_{\perp})\tilde{D}+(d_{\perp}-4)\phi}+V_{\text{RR}}e^{d\tilde{D}+(10-d-2p)\phi}.
\end{align}
For $d=3$ and $d_\perp=2$, this leads to an exponent
\begin{align}
K=1+\frac{2(1-p)}{79+2p(p-9)}.
\end{align}
We find acceleration in the case of massive Type IIA with $F_0$ flux. Since this model is three-dimensional, the NS5 branes are particles in the FRW spacetime. This seems promising for the stability of the network, since if the density $\rho_5$ is small enough, then the particles are widely separated and have no open-string tachyon between them. 

To summarize, we have used brane networks to construct a wide variety of accelerating solutions in critical string theory.  It would be interesting to generalize these models to the supercritical case, as well as to further analyze the stability of the solutions. Also, we have chosen to work in the effective lower-dimensional description here; we leave the full 10-dimensional analysis to future work.

\section{Applications and future directions}\label{sec:applications}

In this paper, we have introduced new classes of string-theoretic models of accelerated expansion.  Our priority has been to obtain explicit examples with a small list of ingredients.  This led us to revisit the many mechanisms for inflation that do not rely on single-field slow roll dynamics in order to generate accelerated expansion, seeking concrete UV complete string-theoretic examples.  In particular, this allows us to construct inflationary solutions from basic exponential potentials in string theory without first metastabilizing the moduli.  So far, this led us to two classes of examples:  in \S\ref{SC}\ a version of assisted inflation for the radii, string coupling, and axions (generalizing and simplifying previous models \cite{SCdS}) and \S\ref{sec:density}\ a UV completion of domain-wall driven acceleration.  
The list of ingredients is relatively small, and includes the leading sources of stress energy in string theory and the dominant axion contribution to the spectrum.  The models of \S\ref{subsec:simpletori}\ and \S\ref{subsec:singleterm}\ are particularly simple.  We expect many more sequence of models along the same lines, for example ones realizing more of the spectrum arising in our warmup toy model \eqref{Kns2}.    

Other mechanisms to translate simple terms in the effective action into accelerating solutions are still in progress \cite{rollingdS}.  One class includes strong warping along one internal spatial direction to produce a de Sitter solution starting from an exponential potential with a tadpole in one higher dimension.   Other approaches incorporate axions, locked inflation at saddle points, or repeated particle production events in different ways.  

We hope these models will prove useful for conceptual and phenomenological applications.  In this section we briefly explore some potential implications, leaving a full treatment for later work.   

\subsection{Inflation, Dark Energy, Axions}

In \S2 we have presented new sequences of models with accelerated expansion, which come along with the large number $\sim 2^D$ of axion fields that dominate the spectrum of string theory.    
Accelerated expansion is well-established in the observed universe, as is dark matter.  The detection of dark energy \cite{SNDE}\ is extremely significant \cite{Huterer}, with contributions from multiple observational probes.  The  detection of a small tilt of the primordial power spectrum \cite{Tilt}\ and other cosmological measurements support the theory of inflation and provide some constraints on its phenomenology.  In fact, power law inflation driven by a single exponential potential is ruled out observationally, so although our models are somewhat more general than that we will focus here on dark energy (it would also be interesting to explore the axion phenomenology of this type of model).    

In this paper, we were led to sequences of models with a variety of equations of state
\beq
w=-1+\frac{2}{3K}
\eeq
with $K$ given by \eqref{Ksimple}\ (\ref{Ktoroidal}) (\ref{Kheavy}).  The sequence \eqref{Ksimple}\ arises in our simplest sequence.  
In the sequence (\ref{Ktoroidal}) there is an accumulation toward $w=-1$, but there is no strict $w=-1$ de Sitter solution with the ingredients contained in the models of \S\ref{subsec:tori}.  

As discussed above in \S2.5, the couplings are weak in our explicit models.  In power law acceleration from the bottom up, interactions of the rolling field $\Phi_\text{c}$ are suppressed at large $K$ (small $\beta_\text{c}$ in the potential $e^{\beta_{\text{c}}\Phi_\text{c}}$).  There is an approximate shift symmetry which protects against large corrections to the potential, such as mass shifts.  In a realistic version of the model, the Standard Model
will generate corrections that depend on the moduli, and this contribution combined with near-canceling fluxes \cite{BP}\ would figure into the assisted inflation mechanism.
One possibility is for the Standard Model to respect the shift symmetry, not coupling directly to $\Phi_\text{c}$.\footnote{This sort of modular structure arises in various scenarios for string-theoretic particle phenomenology.}  Given a realistic version of the mechanism, there is one parameter that is tuned, which can be taken to be the time $t_0$ at which the potential $\propto e^{\beta_\text{c}\Phi_\text{c}}$ is of order our present vacuum energy.  

This is just one sequence of models in what is likely to be a much larger collection, as with previous classes of plausible metastable de Sitter solutions that have been proposed over the years such as \cite{SCdS,KKLT,large,Saltman,Danielson}\ and other top-down examples of quintessence such as \cite{SandipDE}.  Nonetheless it is interesting to contemplate connections to dark energy research.   One key question of observational interest is whether there is any threshold value of  $w$ which distinguishes robust classes of mechanisms, or if there is a preferred value of $w$.  

The value $w=-1$, corresponding to a metastable minimum of the scalar potential, is a special case.  On the other hand, in order to obtain more general equations of state, we consider a simplified set of sources, which may be therefore less generic.  Having done so, however, we generate infinite sequences of possible values of $w<-1/3$.  
With these different possibilities, and the limitations of our current knowledge of model statistics, initial conditions, and other relevant factors, we clearly cannot conclude that $w=-1$ is preferred from the point of view of string theory model statistics or from the point of view of Wilsonian naturalness and fine-tuning, a point also made recently in other top-down models \cite{SandipDE}.   
However, one has to analyze the conditions under which the distinction becomes observationally accessible (see e.g. \cite{KalloshLindeDE,quintessence} for a discussion of this); certainly many values of $w\ne -1$ are observationally degenerate with a pure cosmological constant.  

It is worth emphasizing that neither traditional metastable landscape models, nor the models with more general equations of state, have explicit realizations that are fully realistic. This is simply because both the computation of the Standard Model contribution to the moduli potential and the tuning of microscopic stress-energy sources that nearly cancels it are prohibitively difficult.  In both the cases $w=-1$ and $w\gtrsim -1$, it seems plausible that such tuning is possible and that the resulting moduli potential produces classes of accelerated expansion analogous to those we find in the explicit, but unrealistic, constructions.  But this limitation is another reason that it is not possible (given our current knowledge) to make a robust prediction for $w$ in the string landscape.  As in inflationary cosmology, the next best thing is to analyze different mechanisms for dark energy and distinguish them observationally as far as possible.

\subsection{Holography}

One of our motivations for this work is the prospect of using simpler, more explicit models to help develop a holographic framework for cosmological spacetimes.  In this section, we make some preliminary comments about this application, something we plan to continue in future work.  

As discussed in \cite{dSdS, micromanaging}, inflating solutions admit at least a semi-holographic description, in a way that lines up well with the basic structure of string compactifications.   The infrared regions (those near $w=0, \pi/L$) of the warped metric
\beq\label{dSslice}
ds^2_{dS_{d}}=\sin^2\left(\frac{w}{L}\right) ds^2_{dS_{d-1}}+dw^2
\eeq     
on the de Sitter static patch correspond to a low energy theory which is cut off at a finite scale and coupled to $d-1$ dimensional gravity.\footnote{Similarly in the dS/CFT correspondence \cite{dSCFT}, part of the physics is captured by the dual field theory, but computation of general observables  ultimately involves integration over metrics.}  In other words, de Sitter spacetime itself is a warped compactification, with highly redshifted low-energy regions that admit a dual holographic description.  The explicit models in \cite{micromanaging}\ of large-radius de Sitter spacetime reproduced this general structure, but are still rather complicated to analyze in detail.  

This structure suggests that the dual theory need not be UV complete in itself, since it does not extend all the way to the deep ultraviolet; it may be analogous to theories with a Landau pole at high energies (beyond the scale corresponding to $w=\pi L/2$ in (\ref{dSslice})).  There are existing examples\footnote{see e.g. \cite{UVincomplete,Dong:2012ena}}\ of dual theories which make sense within a warped compactification but which would be unstable or non-unitary if extended to the deep UV.    

In this dS/dS duality framework -- or others such as \cite{dSCFT,FRWCFT}\ -- it should be useful to use explicit models to develop and test these ideas.  We will next make some preliminary comments about this for the two classes of models in this paper.  

\subsubsection{Scaling solutions as in \S\ref{subsec:tori}}  

In flux compactifications such as those we have constructed in 
\S\ref{subsec:tori}, we can get some clues about the putative dual theory in the following way.  
To explain this strategy, let us first review a relevant feature of the duality between the ${\cal N}=4$ super Yang-Mills theory and the type IIB flux compactification on $S^5$.  On the Coulomb branch of this system, the 5-form flux threading the $S^5$ is sourced by explicit domain wall D3-branes on the gravity side.  These branes exhibit a spontaneously broken ${\cal N}=4$ supersymmetric $U(N)$ gauge theory on their worldvolumes, something which might have given a clue about the duality if it had not already been conjectured.    

We can similarly trade the fluxes for branes \cite{TASIlectures}\ in more general flux compactifications, such as those we have developed here, and learn about the content and couplings of the dual theory in a phase in which its scalar fields are turned on.  In the generic case, there will not be an exact moduli space; these fields will be sourced by a potential and hence be time-dependent.  

Implementing this in the models of \S\ref{subsec:tori}\ gives us stacks of D-branes each with a Yang-Mills coupling depending in a very simple way under time evolution:  
\beq\label{gYM}
g_{\text{YM}}\sim g_{\text{YM},0}\frac{t_0}{t}\,.
\eeq
In particular, there are two types of stacks of branes, corresponding to the two types of RR flux described in \S\ref{subsec:tori}.  If we work in the approximation $L_1\approx L_2$ at large $k$ (near de Sitter), these both behave like
\beq\label{gYMagain}
\frac{1}{g_{\text{YM}}^2}\sim e^{-\Phi}e^{\sigma_1} \sim e^{-\Phi}\frac{V_\text{O}}{D}\sim e^{-\Phi} 2^{n/4}\propto\frac{t^2}{t_0^2}\,.
\eeq
At large $K$, i.e. for $w$ approaching $-1$ (near de Sitter), this is very slow evolution relative to the scale factor.  
Power-law time dependent couplings have an interesting RG structure, changing the effective scaling dimension of the couplings and shifting unitarity bounds~\cite{Dong:2012ena,Dong:2012ua}.
In particular, this scaling (\ref{gYM})  implies a classically irrelevant coupling in the $3$-dimensional gauge theory on the branes, a feature that may line up with our general comments above that the dual QFT need not be UV complete in itself given the finite ultraviolet scale in (\ref{dSslice}).
   
From the gravity side, the entropy in the large $K$ (equivalently large $k$) limit of these scaling models behaves as 
\beq\label{SGR}
S\sim \frac{M_4^2}{H^2}\sim \frac{t^2M_4^2}{K^2}\sim \frac{M_4^2}{V_*e^{\Phi(t)}}\,.
\eeq
This simple time-dependence $S\propto t^2$ (for all the scaling solutions) may be related to the simple and universal result above for the time-dependence of $g_{YM}$.  

The dependence on the internal dimensionality $n$ is very interesting as well, with various contributions; in general there is also dependence on the noncompact dimensionality $d = D-n$. The branes carry many bifundamental degrees of freedom, and have a large fermion spectrum from the Ramond sector, with a number of spinor degrees of freedom of order $2^{n/2}$ (as well as a large number of bosonic fields from the NS sector in Neuman-Dirichlet directions). With the explicit models laid out in this paper, we hope to be able to test these ideas for deriving the dual degrees of freedom in detail, using the discrete parameters in the models.  We would like to understand if there is a large-D simplification to the holographic duals and their count of entropy -- both at the cosmological horizon and in D-brane black hole solutions within our new large-radius spacetimes.     

Another interesting consequence of the large dimensionality is that the perturbative expansion parameter in supergravity is weighted by the large suppression factor (\ref{Omegaloop}). Our first class of models in \S \ref{subsec:simpletori} did not use this effect (since $g_s$ was small by itself); however, the models in \S \ref{subsec:tori} approaching the de Sitter limit made use of this suppression factor. It would be very interesting to understand the consequences of having large $g_s$ but a small loop expansion coupling, a point that we hope to analyze in the future. In particular, it will be important to analyze quantum effects on D-brane probes, as well as nonperturbative corrections to the supergravity action.

\subsubsection{Finite density sources as in \S\ref{sec:density}}

Although we had toroidal compactifications in mind in \S\ref{sec:density}, we could consider a generalization that is closer to known holographic models.  In particular, the $d=5$ example in \S\ref{subsec:singleterm}\ could be formulated on an $S^5$ Freund-Rubin compactification of type IIB string theory, with 7-branes wrapping four-spheres within the $S^5$ contributing a domain wall network in the remaining 5 spacetime dimensions.  As long as the density dominates the dynamics (over the $S^5$ curvature and  flux contributions), the system will accelerate as derived above.  It would be interesting to analyze this case more carefully to check this and assess the fate of potential tachyons.

In general, it is very interesting to consider the holographic interpretation of such densities.  First, recall that densities play a role in some string-theoretic AdS/CMT systems such as the Lifshitz theories constructed in \cite{HPST}.  In those systems, the density of branes does not extend all the way to the boundary, since a constant density per unit volume on the gravity side would translate to an infinite density in the dual field theory.  Instead, the density cuts off at a finite radial position, and supports a solution whose dual field theory has Lifshitz scaling in the deep infrared but asymptotes to a conformal field theory in the deep ultraviolet. 
In our present application, the gravity solution is an accelerating cosmology, and as discussed above the dual one infers from the de Sitter static patch does not extend to the deep ultraviolet.  Again, this seems to fit with the fact that a QFT dual to a uniform density on the gravity side must be cut off at a finite scale.  In the Lifshitz example, it crosses over to a different UV field theory, whereas in the present examples there is no additional ultraviolet regime.

\section*{Acknowledgements}

We are grateful to S. Kachru, R. Kallosh, A. Linde, S. McCandlish, J. Polchinski, and T. Wrase for useful discussions.  We would also like to thank the KITP for hospitality during the programs ``Primordial Cosmology" and ``Black Holes:  Complementarity, Fuzz, or Fire?".  
This work was supported in part by the National Science Foundation under grants
PHY-0756174 and NSF PHY11-25915, by the Department of Energy under contract DE-AC03-76SF00515.

\begingroup\raggedright

\endgroup

\end{document}